\documentclass[10pt]{article}
\usepackage[dvips]{graphicx}
\usepackage{amsfonts,verbatim}
\usepackage{amsmath,amssymb}
\usepackage{amsmath,amsbsy}
\usepackage{epsf,epsfig,verbatim}
\usepackage{multirow}
\usepackage{slashbox}
\usepackage{hhline}

\newcommand{\ra}{\rightarrow}
\newcommand{\beq}{\begin{equation}}
\newcommand{\eeq}{\end{equation}}
\newcommand{\C}{{\bf \Large \mathcal{C}}}
\newcommand{\mc}{\mathcal}
\newcommand{\mb}{\mathbb}
\newcommand{\mf}{\mathbf}
\newcommand{\tl}{\tilde}
\newcommand{\h}{\hat}

\newcommand{\e}{\epsilon}

\newcommand{\Ae}{A_\epsilon^{(n)}}

\newtheorem{thm}{Theorem}
\newtheorem{dfn}{Definition}
\newtheorem{rem}{Remark}
\newtheorem{lem}{Lemma}

\allowdisplaybreaks

\begin{document}

\title{A Graph-based Framework for Transmission of Correlated Sources
  over Broadcast Channels 
\footnote{This work was supported by NSF CAREER Grant
  CCF-0448115. This work was presented in part at the 39th Conference
  on Information Sciences and Systems (CISS), Baltimore, MD, March
  2005, Information Theory and Applications workshop 
  (ITA), San Diego, CA, February, 2006, and 
IEEE International Symposium on Information
  Theory (ISIT), Seattle, WA, July 2006.}} 
\author{Suhan Choi and  S.\ Sandeep Pradhan\\
      \{suhanc, pradhanv\}@eecs.umich.edu,\\
EECS Dept., Univ. of  Michigan, Ann Arbor, MI.}

\maketitle

\begin{abstract}
In this paper we consider the communication problem that involves
transmission of correlated sources over broadcast channels. We
consider a graph-based framework for this information transmission
problem. The system involves a source coding module and a channel
coding module. In the source coding module, the sources are
efficiently mapped into a nearly semi-regular bipartite graph, and
in the channel coding module, the edges of this graph are reliably
transmitted over a broadcast channel. We consider nearly
semi-regular bipartite graphs as discrete interface between source
coding and channel coding in this multiterminal setting. We provide
an information-theoretic characterization of (1) the rate of
exponential growth (as a function of the number of channel uses) of
the size of the bipartite graphs whose edges can be reliably
transmitted over a broadcast channel and (2) the rate of exponential
growth (as a function of the number of source samples) of the size
of the bipartite graphs which can reliably represent a pair of
correlated sources to be transmitted over a broadcast channel.
\end{abstract}

\setlength{\baselineskip}{18.6pt}

\section{Introduction}

With the emergence of new set of applications such as wireless sensor
networks, the problem of transmission of correlated information
sources over multiterminal channels has received a renewed attention. 
In this problem, many correlated information sources are accessed by a
set of transmitter terminals, and they wish to simultaneously transmit
some subset of them to another set of receiver terminals over a channel. 
In this paper we address the one-to-many communication system, where 
one transmitter terminal  has access to all the information sources,
and wish to transmit them to many receiver terminals. One such model
involving two receiver terminals was considered by Han and Costa  in 
\cite{han-costa87}, and  is described in the
following. Consider a pair of correlated discrete 
memoryless sources $(S,T)$ with some generic joint distribution
$p(s,t)$ and a pair of finite alphabets $\mc{S}$ and $\mc{T}$,
respectively.  The encoder observes long sequences of realizations of these
sources (of length say $n$),  and wishes to transmit them over a
broadcast channel which has one input $X$ and two outputs $Y_1$ and
$Y_2$, and  Receiver i has access to $Y_i$.  The channel behavior is
governed by a generic conditional distribution $p(y_1,y_2|x)$. The
channel is assumed to be discrete memoryless and is used without
feedback.   The encoder maps $n$-length source sequence pairs into
$n$-length channel input sequences. Each receiver  maps its
corresponding $n$-length channel output sequences into its
corresponding $n$-length source reconstruction sequences. The receivers
would like to produce a reconstruction sequence pair such that the
probability that this deviates from the original pair goes to zero as
blocklength $n$ becomes large. If it is possible to build such
sequences of mappings, then we say that the source pair is
transmissible over the broadcast channel. The goal is to find the set
of all sources that are transmissible over a given broadcast channel
or the set of all channels over which a given source pair is
transmissible. 

Two approaches have been proposed for this problem in the
literature. On is called the joint-source-channel coding approach
and the other is the separation-approach. The former approach
\cite{han-costa87}  addresses the problem directly by finding the
mappings for the given source pair and broadcast channel. A sufficient
condition for transmissibility of a source over a channel
has been given in \cite{han-costa87} for this
problem. In the separation-approach, we divide the encoding task into two
sub-tasks, and similarly the decoding task is accomplished in two
steps. In this approach, the $n$-length source pair is mapped into
three indexes (referred to as $W_0$, $W_1$ and $W_2$) 
coming from three finite sets of size say $\Delta_0$,
$\Delta_1$ and $\Delta_2$, respectively. The first index (common
message) is meant for both receivers, and the second and the third indexes
(private messages)  are meant for Receiver 1 and Receiver 2,
respectively. This is called source encoding. The goal is to remove
all the redundancy from the pair to produce three independent bit
streams. Then these three indexes are mapped (referred to as channel
coding) to $n$-length channel input sequences. On the other side of
the channel, 
 Receiver i first maps its $n$-length channel output
into a pair of indexes ($W_0$ and $W_i$) for $i=1,2$. Then they
independently produce source reconstruction sequences from the common
and the private messages. The first goal is to find the set (called as
the rate region) of all  
the rate tuples $(R_0,R_1,R_2)$, where $R_j=\frac{1}{n}\log \Delta_j$
for $j=0,1,2$, 
at which a  reliable representation of the given source pair can be
accomplished. The second goal is to find the set (called as the capacity
region) of all rate tuples 
$(R_0,R_1,R_2)$, at which a reliable communication of indexes over
the given channel can be accomplished. 
The source coding part works under the assumption that the channel is
noiseless, and the channel coding part works under the assumption that
the messages are independent. 

The channel coding part by itself was first introduced  by Cover
in \cite{cover72,cover75}.
 The capacity region has been found for many interesting classes of
 broadcast channels 
\cite{cover72,cover75,bergmans73,bergmans74,gallager74,
ahlswede-korner75,korner-marton77,elgamal79, 
marton77,pinsker78,marton79,
gelfand80,elgamal}.  The capacity region of certain class of
broadcast channels used in wireless communication systems 
have been obtained in 
 \cite{tse97,li01,caire03,viswanathtsc03,vishwanathjgda03,yu04,weingarten04}.
Marton \cite{marton79} (also see \cite{meulen75,hajek79,elgamal81})
established an inner bound to the capacity
region for the discrete memoryless broadcast channel, which
contains all the known achievable rate regions. 
Outer bounds to the capacity region have been obtained in
\cite{sato78,marton79,nair06}. See \cite{cover98} for a latest 
survey of the results on broadcast channels.
The source coding part was addressed by Gray and Wyner in
\cite{gray74}, where a complete characterization of the rate region 
was given. We refer this source coding problem as Gray-Wyner problem. 
However, it is well-known that the separation-approach is not
optimal for the one-to-many communication problem, unlike the case of
point-to-point information transmission 
problem.  In particular, a simple  example was given in
\cite{han-costa87} that 
showed that a triangular source can be sent reliably over a Blackwell
channel by  using a simple joint-source-channel coding scheme, but
there is no way of 
transmitting that source over that channel using the separation-approach.

Loosely speaking, in the separation-approach, there is an interface
between the source coding module and the channel coding module. 
Due to the structure of the system, i.e.,  a common message and a pair
of private messages, the interface can be thought of as 
a finite collection of products of finite sets. 
For example, in a system with $\Delta_i$ denoting the size of the
$i$th message for $i=0,1,2$, the interface can be thought of as
$\cup_{i=1}^{\Delta_0} (A_i \times B_i) $, where $A_i=\{(i-1)\Delta_1+1, 
(i-1)\Delta_1+2, \ldots, i\Delta_1\}$ and 
$B_i=\{(i-1)\Delta_2+1, 
(i-1)\Delta_2+2, \ldots, i\Delta_2\}$.
However, as seen above, this interface, a finite collection of
products of finite sets, 
is not an efficient representation of a pair of correlated sources for
transmission over broadcast channels\footnote{Another reason why this
  approach may be suboptimal is given in the following. In the
  characterization of the rate region in \cite{gray74}, it turns out
  that for certain choices of the triple $(R_0,R_1,R_2)$ that belongs
  to the boundary of the rate region, i.e., optimal triple, 
the private indexes produced by the joint
  encoder will \emph{not} be independent asymptotically. Hence, the
  channel coding module that follows, which works under the assumptions of
  independence, can not exploit this correlation, and is wasteful of
  resources.   For example, in (15a) in \cite{gray74}, if one chooses
  $W$ such that $(X,W,Y)$ do \emph{not} form a Markov chain, then the triple
$(R_0,R_1,R_2)=(I(X,Y;W),H(X|W),H(Y|W))$ has this property. }.
Note that the corresponding interface for a similar approach in the
point-to-point case is just a finite set. It is also well-known that
a similar separation-approach is not optimal for other multiuser
information transmission problems such as many-to-one communication
\cite{cover-elgamal-salehi80}. 

In our recent work \cite{pradhan-choi-isit,pradhan-choi06}, we have
reported a  bipartite graph-based framework for the problem of
transmission  of information in the many-to-one case. A similar
approach was also studied in \cite{ahlswede-han83}. In the present
work, we consider a similar approach to the one-to-many communication
scenario. The fundamental motivation for this comes from the
concept of typicality \cite{cover-thomas}.  Given a correlated
source pair $(S,T)$, a sequence in $\mc{S}^n$
is said to be typical (or individually  typical) with respect to
$p(s)$, if its empirical 
histogram is close to $p(s)$. Similarly one can define typical
sequences in $\mc{T}^n$. A sequence pair in $\mc{S}^n \times \mc{T}^n$
is said to be jointly typical if its empirical joint histogram is
close to $p(s,t)$. Using the law of large numbers, it follows that 
(a) there are roughly $2^{nH(S)}$ and $2^{nH(T)}$ individually typical
sequences in $\mc{S}^n$ and $\mc{T}^n$, respectively, where $H(\cdot)$
denotes entropy \cite{cover-thomas}, (b) there are roughly
$2^{nH(S,T)}$ jointly typical sequence pairs in $\mc{S}^n \times
\mc{T}^n$, (c) for every typical 
sequence in $\mc{S}^n$, there are roughly $2^{nH(T|S)}$ typical
sequences in $\mc{T}^n$ that are jointly typical and vice versa, 
(d) probability, under $p(s,t)$, of the set of jointly typical sequences
(called jointly typical set)
is close to $1$, and (e) the probability, under $p(s,t)$, of every
jointly typical sequence pair is roughly equal to
$2^{-nH(S,T)}$. These five properties lead one to associate a
bipartite graph on the jointly typical set, with vertexes
formed by individually typical sequences, and two vertexes are connected by
an edge if they are jointly typical. Such bipartite graphs, where the
degrees of the vertexes of one set is close to one constant, and the
degrees of that of the  other set is close to another constant, are referred to
as nearly semi-regular \cite{vanlint}.  Hence graphs can naturally
capture the behavior of the source pair. The details regarding the
source distribution can be dispensed with, and one can just work with
this bipartite graph. This may also lead to the
possibility of using them as discrete interface for one-to-many
communication. The source encoder would now act on the source pair and
produce correlated messages, or edges in a bipartite graph, and the
channel encoder would now work with correlated messages and reliably
transmit the edges in the graph over the broadcast channel. We would still
have a source coding module and a channel coding module. However, now
they would be interfaced using nearly semi-regular bipartite graphs
rather than just a finite collection of products of finite sets. Of
course, a finite collection of products of finite sets is a special
case of nearly semi-regular bipartite graphs.

We now present a brief summary of the results presented in this paper,
for which we need some definitions. A nearly semi-regular bipartite 
graph is said to
have parameters $(\theta_1,\theta_2,\theta'_1,\theta'_2)$ if the $i$th
vertex set has size nearly equal to $\theta_i$ for $i=1,2$, and the
degrees of 
vertexes of the first set is nearly equal to $\theta'_2$ and vice
versa. With a slight abuse of notation, a 
nearly semi-regular bipartite  graph is said to
have parameters $(\theta_0,\theta_1,\theta_2,\theta'_1,\theta'_2)$
if it is the union of $\theta_0$ disjoint subgraphs each having
parameters $(\theta_1,\theta_2,\theta'_1,\theta'_2)$.  
A tuple of rates  $(R_0,R_1,R_2,R'_1,R'_2)$ is said to be achievable
for the given broadcast channel if \emph{there exists} a bipartite graph 
 with parameters $(2^{nR_0},2^{nR_1},2^{nR_2},2^{nR'_1},2^{nR'_2})$
 whose edges can be reliably transmitted by using the channel $n$
 times for large $n$. Similarly, a tuple of rates
 $(R_0,R_1,R_2,R'_1,R'_2)$ is said to be 
achievable for a given pair of correlated sources if \emph{there
  exists} a bipartite graph with parameters
$(2^{nR_0},2^{nR_1},2^{nR_2},2^{nR'_1},2^{nR'_2})$  which can reliably
represent $n$ realizations of the pair for large $n$. 
We provide information-theoretic partial 
 characterizations of the sets of achievable tuples for a 
 broadcast channel  and a correlated source pair. 
These are presented in Theorem 1-4. 
Having the significance of the proposed framework mentioned first, let
us look at the limitations of this optimistic framework as well. 
For that we need to look at the big picture. 

For the point-to-point case, to
check the transmissibility of a source over a channel, we just need to
check the non-emptiness of the intersection of two intervals
$[H,\infty)$ and $[0,C]$, where $H$ denotes the entropy of the source,
  and $C$ denotes the capacity of the channel. Note that only one parameter
  specifies the interval. For the one-to-many communication, the
conventional separation-approach gives a sufficient condition for checking the
  transmissibility: non-emptiness of the rate region of the
  source pair and that of the broadcast channel. We need three
  parameters $(R_0,R_1,R_2)$ to specify the rate region in both source
  coding as well as channel coding.  
Clearly, a  characterization that involves the fewest number of
parameters is what we want.

In the graph-based framework, we consider rate regions for the source
and the channel, which are specified using five parameters:
$(R_0,R_1,R_2,R'_1,R'_2)$.  However, the non-empty intersection of
the rate region of the source pair with that of the channel still does not
guarantee successful transmission. This is because, graphs having the
same set of parameters may have different structures. It turns out
that these
graphs (that have the same set of parameters) can be partitioned into
equivalence classes, where all graphs 
in an equivalence class have the same structure. This structure of the
graphs has also been studied under the name of graph isomorphism
\cite{vanlint} in the literature.  We will address this
issue more formally in later sections.  Hence a graph with
parameters (say) $(2^{nR_0},2^{nR_1},2^{nR_2},2^{nR'_1},2^{nR'_2})$ 
which can reliably represent a source pair may not belong to the
equivalence class of a graph with the same set of parameters whose
edges can be transmitted reliably over the channel. 
In other words, to guarantee successful transmission of the source
over the channel, we need to
construct at least one pair of transmission systems (one for the source
component and one for the channel component) for every equivalence class.
Rather, in this work we have shown (a) the existence of an  equivalence
class for which a transmission system could be built in the source
coding component, and (b)  the existence of an  equivalence
class for which a transmission system could be built in the channel
coding component.
We plan to address this issue further in our future work. 
But we believe that the results given in this paper may be a first
step toward an optimal discrete interface for multiuser
communication. 

The outline of the remaining part of this paper is as follows. In
Section \ref{sec:bc-prelim}, we first provide a summary
of the results in a formal setting that are available in the
literature that are closely related to our work. Then, in 
Section \ref{sec:bc-problem}, we formulate the problem and consider
certain properties of bipartite graphs that are relevant to our
later discussion. Then channel coding part will be discussed in
Section \ref{sec:general-bc}, resulting in an achievable rate region
for the broadcast channel with correlated messages. We also consider
the special case of broadcast channels with one deterministic
component. Thereafter, the complementary source coding part, the representation
of correlated sources into graphs, will be described in
Section \ref{sec:bc-sc}. After that, an example and its
interpretation are provided in Section \ref{sec:bc-example}.
Finally, Section \ref{sec:bc-conclusion} provides some concluding
remarks.

\section{Preliminaries}\label{sec:bc-prelim}

In this section, we provide an overview of the most important prior
results in the literature on broadcast channels, and the related
source coding problem,  which are closely related to our work.

\subsection{Broadcast Channel with Independent private Messages} 
\begin{figure}[!h]
\centering \epsfig{file=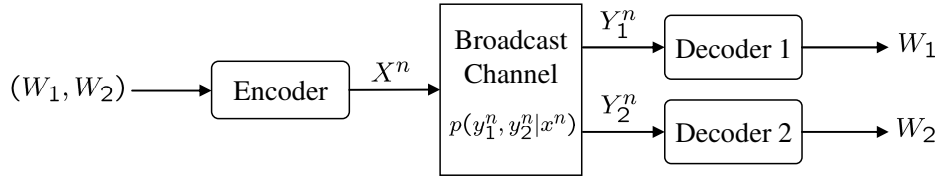, clip=, width=.7\linewidth}
\centering \caption{\small Broadcast channel}\label{fig:broadcast}
\end{figure}

A broadcast channel is composed of one sender and many receivers. 
The objective is to broadcast information from a sender to the many
receivers. We consider broadcast channels with only two receivers
since multiple receivers cases can be similarly treated. Figure
\ref{fig:broadcast} shows a broadcast channel with one sender and
two receivers. The discrete memoryless stationary broadcast channel consists of
an input alphabet $\mathcal{X}$ and two output alphabets
$\mathcal{Y}_1$ and $\mathcal{Y}_2$ and a conditional distribution 
$p(y_1,y_2|x)$. The induced $n$-length conditional distribution is
given by $p(y_1^n,y_2^n|x^n)=\prod_{i=1}^n p(y_{1i},y_{2i}|x_i)$ when used
without feedback. In other words, a broadcast channel is an ordered
tuple $(\mc{X},\mc{Y}_1,\mc{Y}_2, p(y_1,y_2|x))$.
We consider the following definitions for transmission of
independent messages over a broadcast channel.

\begin{dfn}
A transmission system with parameters $(n,\Delta_0,\Delta_1,\Delta_2,\tau)$
for the given broadcast channel $(\mc{X},\mc{Y}_1,$ $\mc{Y}_2,
p(y_1,y_2|x))$ is an ordered tuple $(f,g_1,g_2)$, consisting of one
encoding mapping $f$ and two decoding mappings $g_1$ and $g_2$ where
\begin{itemize}
\item $f: \{1,2,\ldots,\Delta_0\} \times 
\{1,2,\ldots,\Delta_1\} \times \{1,2,\ldots,\Delta_2\}
  \rightarrow \mc{X}^n$, 
\item $g_i: \mc{Y}_i^n \rightarrow \{1,2,\ldots,\Delta_0\} \times 
\{1,2,\ldots,\Delta_i\}$, for $i=1,2$,
\item such that a performance measure given by the average probability
  of error satisfies: 
    \[
      \tau= \sum_{k=1}^{\Delta_0}  \sum_{i=1}^{\Delta_1} \sum_{j=1}^{\Delta_2}
\frac{1}{\Delta_0 \Delta_1 \Delta_2} Pr \left[ (g_1(Y_1^n), g_2(Y_2^n)) \neq
((k,i),(k,j)) | X^n=f(k,i,j) \right].
    \]
\end{itemize}
\end{dfn}

\begin{dfn}
  A rate tuple $(R_0,R_1, R_2)$ is said to be \emph{achievable} for the
  given broadcast channel if for all $\e > 0$, and for sufficiently 
  large $n$, there exists a transmission system as defined above
  satisfying $\frac{1}{n}\log \Delta_i > R_i -\e$ for $i=0,1,2$ and with 
  the average probability of error $\tau < \e$.
\end{dfn}

\begin{dfn}
  The \emph{capacity region} $\mathcal{C}_{BCI}$ of the broadcast
  channel is the set of all achievable rate tuples
  $(R_0,R_1, R_2)$. 
\end{dfn}

The capacity region of general broadcast channels is still not known. But
Marton \cite{marton79} and Gelfand and Pinsker \cite{gelfand80} have
obtained an  achievable rate region for the general discrete
memoryless broadcast channel, 
 which is the largest known inner bound to the capacity region.
An achievable region of the discrete memoryless broadcast channel
\cite{csiszar-korner} is given by all rate tuples $(R_0,R_1,R_2)$
satisfying
  \begin{align}
    R_0 &< \min\{I(Z;Y_1),I(Z;Y_2)\},\\
    R_0 + R_1 &< I(Z,U;Y_1), \\
    R_0 + R_2 &< I(Z,V;Y_2), \\
    R_0 + R_1 + R_2 &< \min\{I(Z;Y_1),I(Z;Y_2)\}+I(U;Y_1|Z)+I(V;Y_2|Z)-I(U;V|Z)
  \end{align}
for some $p(z,u,v,x)$ on ${\mathcal Z} \times $ ${\mathcal U} \times
{\mathcal V} \times {\mathcal X}$, where $Z$, $U$  and $V$ are 
auxiliary random variables with finite alphabets ${\mathcal Z}$,
${\mathcal U}$,  and ${\mathcal V}$, respectively  such
that $(Z,U,V) \rightarrow X \rightarrow (Y_1,Y_2)$ form a Markov
chain. 

\subsection{Gray-Wyner problem}
 
Consider a pair of correlated sources with a joint distribution
$p(s,t)$ and with finite alphabets ${\mathcal S}$ and ${\mathcal
  T}$. The sources are assumed to be stationary and memoryless. These
sources are denoted as an ordered tuple $({\mathcal S},{\mathcal
  T},p(s,t))$. 

\begin{dfn}
A transmission system with parameters
$(n,\Delta_0,\Delta_1,\Delta_2,\tau)$  for representing a pair of
correlated sources $({\mathcal S},{\mathcal T},p(s,t))$ is an ordered
tuple $(f_0,f_1,f_2,g_1,g_2)$ consisting of three encoding mappings
$f_0$, $f_1$ and $f_2$,  and two decoding mappings $g_1$ and $g_2$ where 
\begin{itemize}
\item $f_i: \mc{S}^n \times \mc{T}^n \rightarrow 
 \{1,2,\ldots,\Delta_i\},  \ \ \mbox{for } \ \ i=0,1,2$,
\item $g_1:  \{1,2,\ldots,\Delta_0\} \times \{1,2,\ldots,\Delta_1\}
\rightarrow  \mc{S}^n$, \ \ \ 
 $g_2:  \{1,2,\ldots,\Delta_0\} \times \{1,2,\ldots,\Delta_2\}
\rightarrow  \mc{T}^n,$
\item such that a performance measure given by the probability
  of error satisfies: 
    \[
      \tau= Pr \left[(g_1(f_0(S^n,T^n),f_1(S^n,T^n)),
g_2(f_0(S^n,T^n),f_2(S^n,T^n)) \neq (S^n,T^n)\right].
    \]
\end{itemize}
\end{dfn}

\begin{dfn}
  A rate tuple $(R_0,R_1, R_2)$ is said to be \emph{achievable} for the
  given correlated sources if for all $\e > 0$, and for sufficiently 
  large $n$, there exists a transmission system as defined above
  satisfying $\frac{1}{n}\log \Delta_i < R_i +\e$ for $i=0,1,2$, and with 
  the average probability of error $\tau < \e$.
\end{dfn}

\begin{dfn}
  The \emph{achievable region} $\mathcal{R}_{GWI}$ for the correlated
  sources is the set of all achievable rate tuples.
\end{dfn}

An information-theoretic characterization of this rate region
\cite{gray74,wyner75} is given in the
following. For a given pair of correlated sources a  tuple
$(R_0,R_1,R_2)$ is achievable  if and only if 
 \begin{align}
    R_0 &> I(S,T;Z), \\
    R_1 &> H(S|Z), \\
    R_2 &> H(T|Z), 
    \end{align}
for some distribution
$p(z,s,t)=p(s,t)p(z|s,t)$, where $Z$ is an auxiliary random variable with a
finite alphabet ${\mathcal Z}$. Using convexity arguments, it can be
shown that there is no loss of optimality if $|{\mathcal Z}| \leq
|{\mathcal S}||{\mathcal T}|$. 

This problem was  also considered in \cite{wyner75} in a slightly different
form. The minimum $R_0$ that belongs to the rate region such that the
corresponding $Z$ satisfies $T \rightarrow Z \rightarrow S$ a Markov
chain,  is called as
Wyner's common information $C(S,T)$. 

\subsection{Joint source-channel coding}
Consider the joint source-channel coding scheme studied in
\cite{han-costa87}. Suppose we are given a pair of correlated sources
without common part \cite{gacs,wit,wyner75}
\footnote{\cite{han-costa87} considered the general
  setting where sources may have non-zero common part.}   and a
broadcast channel. 

\begin{dfn}A transmission system with parameters $(n,\tau)$ for
  transmission of a pair of correlated sources $({\mathcal
  S},{\mathcal T}, p(s,t))$ and a broadcast channel 
$({\mathcal X},{\mathcal Y}_1,{\mathcal Y}_2,p(y_1,y_2|x))$ is an 
ordered tuple $(f,g_1,g_2)$ where
\begin{itemize}
\item $f:{\mathcal S}^n \times {\mathcal T}^n \rightarrow {\mathcal
  X}^n$ 
\item $g_1: {\mathcal Y}_1^n \rightarrow {\mathcal S}^n$, and 
$g_2: {\mathcal Y}_2^n \rightarrow {\mathcal T}^n$.
\item such that a performance measure given by the probability of
  decoding error satisfies
\[
\tau= \sum_{(s^n,t^n) \in {\mathcal S}^n \times {\mathcal T}^n}
p^n(s^n,t^n) Pr[(g_1(Y_1^n),g_2(Y_2^n))\neq (s^n,t^n)|X^n=f(s^n,t^n)].
\]
\end{itemize}

\end{dfn}

\begin{dfn}
A pair of correlated sources is said to be transmissible over a
broadcast channel if $\forall \e>0$, and for all sufficiently large
$n$, there exists a transmission system as defined above with
parameters $(n,\tau)$ such that $\tau < \e$.
\end{dfn}

The result of \cite{han-costa87} says that a pair of correlated
sources is transmissible over a broadcast channel if 
\begin{align}
H(S) &< I(S,W,U;Y_1)-I(T;W,U|S) \\
H(T) &< I(T,W,V;Y_2)-I(S;W,V|T) \\
H(S,T) &< \min \{I(W;Y_1),I(W;Y_2)\} +I(S,U;Y_1|W)+
I(T,V;Y_2|W)-I(S,U;T,V|W)
\end{align}
for some $p(w,u,v,x|s,t)$ on 
${\mathcal W} \times {\mathcal U} \times {\mathcal V} \times {\mathcal
  X} \times  {\mathcal S} \times {\mathcal T}$, where $W,U,V$ are
auxiliary random variables with finite alphabets 
${\mathcal W}$, ${\mathcal U}$ and ${\mathcal V}$ such that 
$(S,T) \rightarrow (WUV)\rightarrow X \rightarrow (Y_1,Y_2)$ form
a Markov chain. 

This problem has also been considered in different settings recently in
\cite{tuncel05,coleman06}.

\subsection{An Example of transmission of Correlated Sources over the Broadcast
  Channel}\label{sec:bc-ex} 

Let us consider an interesting example given in \cite{han-costa87}
showing the advantage of encoders that exploit the correlation
between sources. Consider the transmission of a set of correlated
sources $(S,T)$ with the joint distribution $p(s,t)$ given by
\begin{center}

\renewcommand{\arraystretch}{1.5}

\setlength{\fboxsep}{2pt}\setlength{\tabcolsep}{3pt}

\begin{tabular}{|c|c|c|}
\hline
 \setlength{\backslashbox{$T$}{$S$}} &0 & 1 \\ \hline
 0 & $~~~\frac{1}{3}~~~$ & 0 \\ \hline
 1 & $\frac{1}{3}$ & $~~~\frac{1}{3}~~~$ \\\hline
\end{tabular}
\end{center}
with finite alphabet $\mc{S}=\mc{T}=\{0,1\}$ over a Blackwell
channel with $\mc{X}=\{1,2,3\}$, $\mc{Y}_1=\mc{Y}_2=\{0,1\}$ where
the channel transition probabilities are specified by
$p(0,0|1)=p(0,1|2)=p(1,1|3)=1$. If we assign $X$=1, 2, and 3 to
$(S,T)$ = (0,0), (0,1), and (1,1), respectively, then $Y_1$ and
$Y_2$ determine $S$ and $T$ without error, respectively.

In the conventional separation-approach, 
first, factor $(S^n,T^n)$ into three independent messages
  $W_1,W_2$, and $W_0$.
Next, transmit $(W_0,W_1)$ and $(W_0,W_2)$ to receivers 1 and 2,
  respectively, over the broadcast channel using some, hitherto unknown,
  optimal coding scheme, so that $(W_0,W_1)$ and $(W_0,W_2)$ reliably
  determine $S^n$ and $T^n$, respectively.
Let the rate of $W_i$ be $R_i=\frac{1}{n}H(W_i)$ for
$i$ = 0, 1 and 2. Then, as the channel in consideration is
deterministic, the sum of these rates must be bounded as  $R_0
+ R_1 + R_2 \leq \frac{1}{n} H(Y_1^n, Y_2^n) \leq \frac{1}{n} H(X^n)
\leq \log_2 3=H(S,T)$.  

According to \cite{wyner75}, the most efficient
decomposition of this kind with the constraint $R_0 + R_1 + R_2
\leq H(S,T)$ is attained when $R_0=C(S,T)$ In this case,
\begin{align}
  R_0 + R_1 + R_2 &= H(S,T) = \log_2 3 \textrm{(bits)}\\
  2R_0 + R_1 + R_2 &= \log_2 3 + I(S,T;Z) \geq  \log_2 3+C(S;T).
\end{align}
For this triangular source,
$C(S;T)=\log_2 3-h_2(\frac{2}{3}) = \frac{2}{3}$ where
$h_2(p)=-p\log_2 p-(1-p)\log_2(1-p)$. Therefore, $2R_0 + R_1 + R_2
\geq \log_2 3 +\frac{2}{3}\approx 2.252$ (bits). Consequently,
$R_0+R_1 > 1$ or $R_0+R_2 >1$. Since $H(Y_1) \leq 1$ and $H(Y_2)
\leq 1$ for any distribution, receiver 1 cannot reliably reproduce
$(W_0,W_1)$ or receiver 2 cannot reliably reproduce $(W_0,W_2)$.
Thus there is no way of reliably transmitting this triangular
source via the Blackwell channel by factoring the sources into $W_1,W_2$, and
$W_0$ as shown above.

\section{Problem Formulation}\label{sec:bc-problem}

\begin{figure}[h]
\centering \epsfig{file=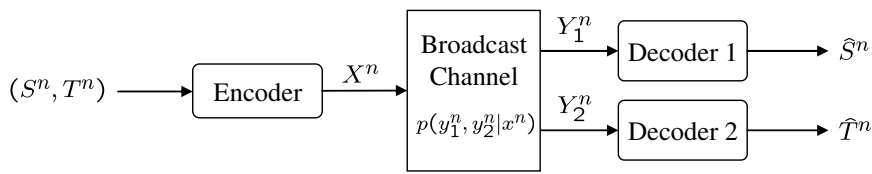, clip, width=0.65\linewidth}
\centering \caption{\small Transmission of correlated sources over a
broadcast channel} \label{fig:bc-block}
\end{figure}

The problem we are addressing is the simultaneous transmission of
two correlated sources $(\mc{S},\mc{T},p(s,t))$ over a
broadcast channel 
$({\mathcal X},{\mathcal Y}_1,{\mathcal Y}_2,p(y_1,y_2|x))$, 
with one sender and
two receivers as shown in Figure \ref{fig:bc-block}. Here, the
encoder can access both sources $(S,T)$ and the receivers can not
communicate with each other. 
The encoder is given by a mapping $f: \mc{S}^n \times \mc{T}^n \ra
\mc{X}^n$. The decoders are given by mappings $g_1:\mc{Y}_1^n \ra
\mc{S}^n$ and $g_2:\mc{Y}_2^n \ra \mc{T}^n$. The performance measure
associated with this transmission system is the probability of
decoding error:
\begin{equation}
  Pr[(S^n,T^n)\neq(g_1(Y_1^n),g_2(Y_2^n))].
\end{equation}

\subsection{Basic Concepts}

\begin{figure}[h]
\centering \epsfig{file=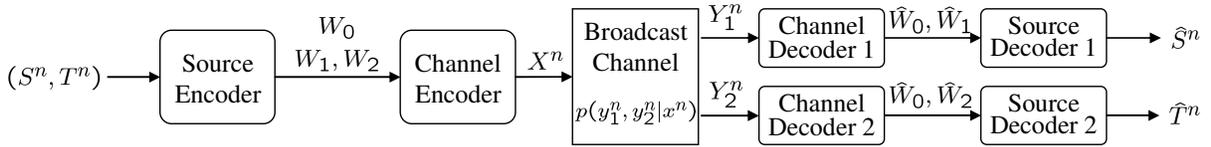, clip,
width=0.9\linewidth} \centering \caption{\small The outputs of
sources are first mapped into edges in a bipartite graph, and the
edges coming from this graph are reliably transmitted over a
broadcast channel} \label{fig:bc-block-sc-cc}
\end{figure}

We consider a modular approach to this problem, which is shown in
Figure \ref{fig:bc-block-sc-cc}. The system has two modules: the
source coding module and the channel coding module. The outputs of
two correlated sources are first represented efficiently into a
triple of messages $(W_0,W_1,W_2)$ in the source coding module.
Then, these messages are reliably transmitted over the broadcast
channel in the channel coding module. 
In more detail, the source coding module produces three
messages $W_0, W_1$ and $W_2$, where $W_0$ is a common message to
both receivers which contains common information about both sources
$S$ and $T$, and $W_1$ and $W_2$ are private messages which contain
the individually remaining information about the source $S$ and $T$
after extracting the common information, respectively. The messages
$W_0,W_1$ and $W_2$ belong to integer sets
$\mathcal{W}_0=\{1, 2, \ldots, \Delta_0\}$, $\mathcal{W}_1=\{1, 2,
\ldots, \Delta_1\}$ and $\mathcal{W}_2=\{1, 2, \ldots, \Delta_2\}$,
respectively. In general, $W_0,W_1$ and $W_2$ are not independent.
Then, the channel
coding module wants to reliably transmit a message pair $(W_0,W_1)$
to receiver 1 and $(W_0,W_2)$ to receiver 2.

We assume that there is some kind of correlation between two
messages $W_1$ and $W_2$, i.e., private messages
 for the receivers can not be chosen independently. 
\begin{dfn}
\begin{itemize}
  \item The private messages of receivers are said to be \emph{correlated}, if
for every $w_0 \in \mc{W}_0$,   there exists a set $A(w_0)$ such that $A(w_0)
\subset \mathcal{W}_1 \times 
\mathcal{W}_2$,  and conditioned on $W_0=w_0$ the message
  pairs $(W_1,W_2) \in A(w_0)$ are equally likely with probability
  $\frac{1}{|A(w_0)|}$, and the message pairs $(W_1,W_2) \notin A(w_0)$ have
  probability zero.
  \item The private messages of the receivers are said to be
  \emph{independent}, 
  if $A(w_0) = \mathcal{W}_1 \times
\mathcal{W}_2$ for all $w_0 \in \mc{W}_0$.  In this case, the message pairs
    $(W_1,W_2)$ are equally likely with probability
    $\frac{1}{|\mathcal{W}_1
\times \mathcal{W}_2|}$.
\end{itemize}
\end{dfn}
We use bipartite graphs to model the correlation of the messages,
i.e., the set $A(w_0)$ is taken to be a bipartite graph for all $w_0
\in \mc{W}_0$. Let us first
define a bipartite graph and related mathematical terms before we
discuss the main problem.

\begin{dfn}
\begin{itemize}
  \item A \emph{bipartite graph} $G$ is defined as an ordered tuple
$G=(A_1,A_2,B)$ where $A_1$ and $A_2$ are two
  non-empty sets of vertexes, and $B$ is a set of edges where every edge
   of $B$ joins a vertex in $A_1$ to a vertex in $A_2$, i.e., $B \subseteq A_1 \times A_2$.
  \item If $G$ is a bipartite graph, let $V_1(G)$ and $V_2(G)$ denote  the
    first and the second vertex sets of $G$, respectively,  and
  $E(G)$ denote the edge set of $G$.
  \item If $(i,j) \in E(G)$, then $i$ and $j$ are said to be
  \emph{adjacent}.
  \item If each vertex in one set is adjacent to every vertex in the other
  set, then $G$ is said to be a \emph{complete} bipartite graph. In this case, $E(G) = V_1(G) \times
  V_2(G)$.
  \item The \emph{degree} of a
vertex $v \in V_i(G)$ in a graph $G$, denoted by
$\mathrm{deg}_{G,i}(v)$, is the number of edges
connected to $v$ for $i=1,2$.
  \item A \emph{subgraph} of a graph $G$ is a graph whose vertex and
  edge sets are subsets of those of $G$.
\end{itemize}
\end{dfn}
Since we consider a specific type of bipartite graphs in our
discussion, let us define those bipartite graphs as well. Although our main
results deal with nearly semi-regular graphs, for the purpose of
illustration, we consider semi-regular graphs for this section
alone.

\begin{dfn}\label{def:graph-4-parameters}
A bipartite graph $G$ is called \emph{semi-regular} \cite{vanlint}
 with parameters $(\theta_1, \theta_2, \theta_1',
\theta_2')$, if it satisfies:
\begin{itemize}
  \item $|V_i(G)|=\theta_i$ for $i$=1, 2,
  \item $\forall u \in V_1(G)$, $\mathrm{deg}_{G,1}(u)= \theta_2'$,
  \item $\forall v \in V_2(G)$, $\mathrm{deg}_{G,2}(v)= \theta_1'$.
\end{itemize}
\end{dfn}

\begin{figure}[h]
\centering \epsfig{file=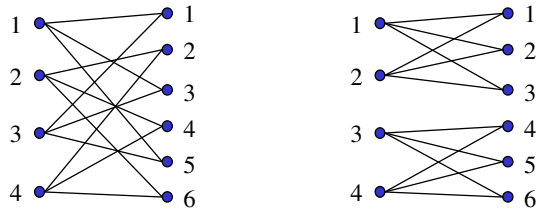, clip,
width=0.4\linewidth} \centering \caption{\small Examples of
semi-regular bipartite graphs with parameters $(4, 6, 2, 3)$}
\label{fig:semiregular-graph-ex}
\end{figure}

As an example, two semi-regular bipartite graphs with parameters
$(4,6,2,3)$ are shown in Figure \ref{fig:semiregular-graph-ex}. Note
that there exist many semi-regular bipartite graphs with the same
set of parameters.

\begin{dfn}\label{def:graph-5-parameters}
 A bipartite graph $G$ is called \emph{nearly semi-regular} with parameters $(\Delta_1$, $\Delta_2$, $\Delta_1'$,
$\Delta_2'$, $\mu)$ for $\mu > 1$ if it satisfies:
\begin{itemize}
  \item $|V_i(G)|=\Delta_i $ for $i=1, 2$,
  \item $\forall u \in V_1(G)$, $\Delta_2' \mu^{-1} \leq
  \mathrm{deg}_{G,1}(u) \leq \Delta_2' \mu$, \ \ \ and \ \
  $\forall v \in V_2(G)$, $\Delta_1' \mu^{-1}  \leq \mathrm{deg}_{G,2}(v)  \leq \Delta_1' \mu$.
\end{itemize}
\end{dfn}
Note that $\mu$ is a slack parameter, which determines the range of
degrees of vertexes.

\begin{dfn}\label{def:bc-graph-5-parameters}
 A nearly semi-regular
bipartite graph $G$ is said to have parameters $(\Delta_0, \Delta_1,
\Delta_2, \Delta_1', \Delta_2', \mu)$ for $\mu > 1$, if it
satisfies the following conditions:
\begin{itemize}
  \item $G$ is the union  of $\Delta_0$ disjoint subgraphs
  $\tilde{G}_m$ for $m=1,2,\ldots,\Delta_0$ where $\tilde{G}_m$ is
  nearly semi-regular with parameters 
$(\Delta_1, \Delta_2, \Delta_1', \Delta_2', \mu)$ as in Definition 
\ref{def:graph-5-parameters}.
  \item $\forall m \in \{1,2,\ldots, \Delta_0\}$,
    $V_i(\tilde{G}_m)=\{(m-1)\Delta_i+1,(m-1)\Delta_i+2,\ldots,m\Delta_i\}$
    for 
    $i$=1, 2,
  \item $V_i(G)=\bigcup_{m=1}^{\Delta_0} V_i(\tilde{G}_m)$ for $i$=1, 2,
  \ \ and \ \  $E(G)=\bigcup_{m=1}^{\Delta_0} E(\tilde{G}_m)$,
 \end{itemize}
\end{dfn}

\begin{figure}[h]
\centering \epsfig{file=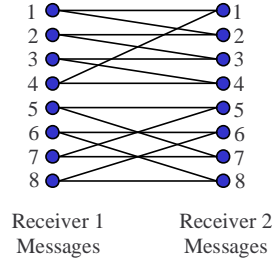, clip,
width=0.2\linewidth} \centering \caption{\small An example of
bipartite graph $G$, with parameters $(2, 4, 4, 2, 2,1)$, composed of
two disjoint  subgraphs $\tl{G}_1$ and $\tl{G}_2$ where $\mc{W}_0=\{1,2\}$,
$\mc{W}_1=\mc{W}_2=\{1,2,3,4\}$.} \label{fig:bc-msg-graph-ex}
\end{figure}

A triple of messages (random variables) $(W_0,W_1,W_2)$ can be
associated with a graph $G$ with parameters $(\Delta_0$, $\Delta_1$,
$\Delta_2$, $\Delta_1'$, $\Delta_2'$, $\mu)$ in the following way.
$W_i \in
\mc{W}_i$ and $\mc{W}_i=\{1,2,\ldots,\Delta_i\}$ for $i=0,1,2$, and
an edge $((m-1)\Delta_1+i,(m-1)\Delta_2+j)\in E(G)$ denotes a
realization of the triple
of messages $(W_0,W_1,W_2)=(m,i,j)$. In other words, 
$Pr[(W_0,W_1,W_2)=(m,i,j)]=\frac{1}{|E(G)|}$ if 
$((m-1)\Delta_1+i,(m-1)\Delta_2+j)\in E(G)$ and $0$ otherwise.
So, different correlation
structures of the messages can be modeled by varying the 
structure of the graph. Figure \ref{fig:bc-msg-graph-ex} illustrates
an example of a bipartite graph $G$ with parameters 
$(2, 4, 4, 2, 2, 1)$ composed of two
disjoint subgraphs. This graph can be associated with a triple of
messages $(W_0,W_1,W_2)$ where $W_i \in \mc{W}_i$ for $i=1,2,3$ such
that $\mc{W}_0=\{1,2\}$, and $\mc{W}_1=\mc{W}_2=\{1,2,3,4\}$. Note
that in the figure two subgraphs $\tl{G}_1$ and $\tl{G}_2$ have
different edge structures. This implies that the correlation of
$(W_1,W_2)$ depends on $W_0$. $W_0$ and $W_1$ are independent, 
and so are $W_0$ and $W_2$.

\subsection{Equivalence classes of graphs}\label{sec:bc-equivalence-classes}

Let us consider the set of all semi-regular bipartite
graphs with fixed parameters
$(\theta_1,\theta_2,\theta_1',\theta_2')$. It is well-known that 
this set can
be partitioned into equivalence classes where equivalence relation is
permutation and relabeling of the vertexes in the graphs. In more
detail, one element (or graph) in a class can be obtained from the
other in the same class by permutation and relabeling of the
vertexes. However, if two elements (or graphs) belong to different
classes, they can not be obtained from each other by permutation and
relabeling since they have different correlation structures.

This means that if we have a transmission system which can reliably
transmit the edges of a graph $G$ (i.e., the correlation of the
message pairs are modeled using $G$) then this transmission system can
be used to reliably transmit edges of any graph  that belongs to the
equivalence class of $G$. Similarly, if one can construct a source
representation system that can represent a source pair using a graph
$G$ (i.e., the correlation of the index pairs produced by the source
encoder is modeled using $G$), then it can be used to represent the
source pair using any graph that belongs to the equivalence class of
$G$.

\section{Broadcast Channels with Correlated
Messages}\label{sec:general-bc}

In this section we characterize transmissibility of certain
correlated messages over a stationary discrete memoryless broadcast
channel.

\subsection{Summary of Results}

\begin{figure}[h]
\centering \epsfig{file=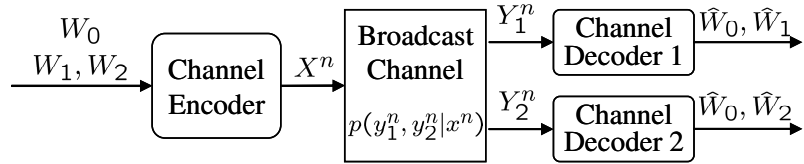, clip=, width=.6\linewidth}
\centering \caption{\small Transmission of correlated messages over
the broadcast channel} \label{fig:bc-cc-block}
\end{figure}

Although, ideally, we would want to use semi-regular graphs for
source representation and communication of information over broadcast
channels, for the sake
of analytical tractability, as is typical in Shannon theory, we will
allow some slack with regard to the degrees of the vertexes of these
graphs, and consider the asymptotic case when this slack is bounded in
some way.

\begin{dfn}\label{def:bc-cc-tx-system}
An $(n,\tau)$-transmission system for a nearly semi-regular
bipartite graph $G$ with parameters $(\Delta_0$, $\Delta_1$,
$\Delta_2$, $\Delta_1'$, $\Delta_2'$, $\mu)$ and a broadcast channel
$(\mc{X},\mc{Y}_1, \mc{Y}_2,p(y_1,y_2|x))$ is an ordered tuple
$(f,g_1,g_2)$, consisting of one encoding mapping $f$ and two
decoding mappings $g_1$ and $g_2$ where
\begin{itemize}
\item $f: E(G) \ra \mc{X}^n$, i.e,\\ $\forall ((m-1)\Delta_1+i,(m-1)\Delta_2+j) \in E(G)$, assign
  $x^n=f(m,i,j)$ where $m \in \{1,2,\ldots,\Delta_0\}$, $i \in \{1,2,\ldots,\Delta_1\}$, and $j \in
  \{1,2,\ldots,\Delta_2\}$,
\item $g_i: \mc{Y}_i^n \ra V_i(G)$ for $i=1,2$, i.e.,
  $g_i: ~\mc{Y}_i^n \ra
  \{1,2,\ldots,\Delta_0\}\times\{1,2,\ldots,\Delta_i\}$,
\item such that a performance measure given by the following average
probability of error satisfies: \beq \tau= \frac{1}{|E(G)|} \sum_{
((m-1)\Delta_1+i,(m-1)\Delta_2+j) \in E(G)} \!\! Pr \left[
(g_1(Y_1^n),g_2(Y_2^n))\!\!\neq\!\! ((m,i),(m,j))
|X^n=f(m,i,j)\right]. \eeq
\end{itemize}
\end{dfn}

In Definition \ref{def:bc-cc-tx-system}, the messages $W_0$, $W_1$
and $W_2$ are assumed to have the following distribution:
\begin{itemize}
  \item Alphabet of $W_i$ is $\{1,2,\ldots,\Delta_i\}$ for $i=0,1,2$,
  \item $Pr\{(W_0,W_1,W_2)=(m,i,j)\}=\left\{
                                       \begin{array}{cl}
                                         \frac{1}{E(G)}, & \hbox{if $((m-1)\Delta_1+i,(m-1)\Delta_2+j)\in E(G)$,} \\
                                         0, & \hbox{else.}
                                       \end{array}
                                     \right.$

\end{itemize}

\noindent
\begin{rem}
\emph{In broadcast channels, the goal of the channel encoder is
to reliably transmit two pairs of messages $(W_0,W_1)$ and
$(W_0,W_2)$ to receiver 1 and receiver 2, respectively, over the
channel. In terms of graphs, it is to reliably transmit edges of a
graph which is associated with the triple of messages
$(W_0,W_1,W_2)$ over the channel.
 As an analogy, the conventional Shannon's
channel coding theorem in a point-to-point communication scenario
can be interpreted as finding the maximum number of codewords
(colors, if each codeword has a different color) that are
distinguishable at the noisy receiver. In the conventional broadcast
channel with one sender and two receivers, the goal is to
distinguish colors at the noisy receivers, where the first color,
which is common to two receivers, can come from one set and the
second and the third colors, which are private to each receiver, can
come from other two sets, respectively, and all possible combination
of triples in the three sets are allowed. A natural question to ask
is: if only a fraction of all possible combination of pairs of
colors is permitted depending on the common color, what is the
maximum size of the sets of these colors which can be reliably
distinguished at the receivers.}
\end{rem}

\begin{dfn}
  A tuple of rates $(R_0,R_1,R_2,R_1',R_2')$ is said to be
  \emph{achievable} for a given broadcast channel with 
correlated messages, if for all $\e > 0$, and for sufficiently
large $n$,
  there exists a bipartite graph $G$ with parameters $(\Delta_0,
  \Delta_1, \Delta_2, \Delta_1', \Delta_2',\mu)$ and an associated 
  $(n,\tau)$-transmission system as defined above satisfying:
$\frac{1}{n}\log \Delta_0 > R_0-\e$,  $\frac{1}{n}\log \Delta_i >
R_i-\e$, $\frac{1}{n}\log \Delta_i' > R_i'-\e$ for $i=1, 2$,
$\frac{1}{n}\log \mu < \e$ and the corresponding average probability
of error $\tau < \e$.
\end{dfn}

Note that in the above definition, we have taken an optimistic point
of view. As long as one can find a sequence of nearly semi-regular
graphs where the number of vertexes and the degrees are increasing
exponentially with given rates, such that the edges from these
graphs are reliably transmitted over the given broadcast channel, we
allow the corresponding rate tuple to belong to the achievable rate
region. The goal is to find the capacity region $\mc{C}_{BC}$ which
is the set of all achievable tuple of rates $(R_0, R_1, R_2, R_1',
R_2')$. In the following we provide an information-theoretic
characterization of an achievable rate region. This is an inner
bound to the capacity region ${\mathcal C_{BC}}$, and is also a
per-letter characterization. This is one of the main results of this
paper. 

\begin{thm}\label{thm:rate-region-bc-cor}
For a discrete memoryless broadcast channel $(\mc{X}, \mc{Y}_1,
\mc{Y}_2, p(y_1,y_2|x))$, $\mc{R}_{BC}^* \subset \mc{C}_{BC}$ where
  \begin{align}
\mc{R}_{BC}^* =\bigcup_{p(z,u,v,x)}\{(R_0, R_1, R_2, R_1', R_2'):~~~& \notag\\
R_0 &\leq \min\{I(Z;Y_1),I(Z;Y_2)\},\\
    R_1 &\leq I(U;Y_1|Z), \\
    R_2 &\leq I(V;Y_2|Z), \\
    R_1 + R_2' = R_1' + R_2& \leq I(U;Y_1|Z)+I(V;Y_2|Z)-I(U;V|Z)\}
  \end{align}
where $Z$, $U$ and $V$ are auxiliary random variables with finite
alphabets ${\mathcal Z}$, ${\mathcal U}$ and ${\mathcal V}$,
respectively, and  
$p(z,u,v,x,y_1,y_2)=p(z)p(u,v|z)p(x|z,u,v)p(y_1,y_2|x)$
satisfies a Markov chain $(Z,U,V) \ra X \ra (Y_1,Y_2)$.
\end{thm}

\begin{rem}
\emph{
When the private messages are
  independent, i.e., when all the elements in the set $\mc{W}_1 \times 
  \mc{W}_2$ can occur with non-zero and equal probability, the rate region
  becomes exactly the same as Marton's  \cite{marton79}. 
  However, when the private messages are correlated, i.e., only some
  elements in the set $\mc{W}_1 \times 
  \mc{W}_2$ can occur equally likely, the sum rate $R_1+R_2$ can be
  larger. As the amount of correlation between the messages increases
  the achievable rate region also becomes larger.  }
\end{rem}

\begin{rem}
\emph{
  The limitations of this theorem are as follows. Note that this
  theorem gives only a partial characterization of the set of all
  nearly semi-regular graphs whose edges can be reliably transmitted
  over a broadcast channel. In the formulation of the achievable rate
  region, we 
  have the freedom of choosing the correlation of the messages for
  every block-length $n$. The theorem characterizes the rate of
  exponential growth (as a function of the number of channel uses) of 
  size of certain nearly semi-regular graphs, such that edges
  coming from any such graph can be reliably transmitted over the
  broadcast channel. 
   This obviously also means that it is possible to transmit
  edges of every graph that belongs to the equivalence class of any of
  these graphs. However, this fact does not mean that the edges of any
  graph with those parameters can be reliably transmitted.}
\end{rem}

\subsection{Proof of Theorem
{\ref{thm:rate-region-bc-cor}}}\label{sec:proof-general-bc}

In this section we prove Theorem \ref{thm:rate-region-bc-cor} by
using the method of random coding, random binning, the concept of joint
typicality of sequence pairs and some concepts from the theory of
random graphs \cite{jansbook}.  In addition to using the
techniques given in \cite{elgamal81}, we devise a concept of a
``\emph{super-bin}'', which is a group of consecutive bins, to take
into account the correlation between the messages.
Given a broadcast channel with distribution $p(y_1,y_2|x)$,
consider a fixed joint distribution $p(z,u,v,x)$ = $p(z)$
$p(u,v|z)$ $p(x|z,u,v)$ where $z,u$ and $v$ are auxiliary random
variables on $\mc{Z} \times \mc{U} \times \mc{V}$. Also, fix $\e
>0$, an integer $n \geq 1$, and positive real numbers $R_0$,
$R_1$, and $R_2$.

\vspace{10pt}

\noindent \textbf{Random sequences and bin generation}: Draw
$2^{nR_0}$ sequences $Z^n(m)$ for  $m \in \{1,2,\ldots,2^{nR_0} \}$, of
length $n$, independently with replacement from $\Ae(Z)$ each with probability
$\frac{1}{|\Ae(Z)|}$ where $\Ae(Z)$ is the strongly $\e$-typical
set with respect to the distribution $p(z)$ which is a marginal of
the joint distribution $p(z,u,v)$ and $|A|$ denotes the
cardinality of a set $A$.

For each $m \in \{1,2,\ldots, 2^{nR_0}\}$, draw
$2^{n(I(U;Y_1|Z)-\e)}$ $n$-length sequences  
$U^n(k,m)$ for $k \in \{1,2,\ldots, 2^{n(I(U;Y_1|Z)-\e)}\}$
independently with replacement from  $A(U|Z^n(m))$ with probability
$\frac{1}{|A(U|Z^n(m))|}$. Call this collection as $\C_1(m)$.
Here, $A(U|z^n)$ is the set of
$n$-sequences $u^n$ which are strongly jointly typical with the
sequence $z^n$, i.e., for $z^n \in A_{\epsilon}^{(n)}(Z)$, the set 
$A(U|z^n) \triangleq \{u^n|(u^n,z^n)\in \Ae(U,Z)\}$.
Similarly, generate $2^{n(I(V;Y_2|Z)-\e)}$ 
sequences $V^n(l,m)$ for $l \in$  $\{1,2,\ldots,$ $2^{n(I(V;Y_2|Z)-\e)}\}$ 
independently with replacement from  $A(V|Z^n(m))$, where 
for $z^n \in A_{\epsilon}^{(n)}$, the set  $A(V|z^n) \triangleq
\{v^n|(v^n,z^n)\in \Ae(V,Z)\}$. Call this collection as $\C_2(m)$.
Without loss of generality $2^{n(I(U;Y_1|Z)-\e)}$ and
$2^{n(I(V;Y_2|Z)-\e)}$ are assumed to be integers.

Next, for each $m$ in $\{1,2,\ldots, 2^{nR_0}\}$, define random bins
$B(i,m)$ and $C(j,m)$ for 
$i \in \{1,2,\ldots, 2^{nR_1} \}$ and $j \in \{1,2,\ldots,
2^{nR_2} \}$ such that
\begin{align}
B(i,m) &= \{U^n(k,m)~|~ k \in [(i-1)\cdot 2^{n(I(U;Y_1|Z)-R_1-\e)}+1,
  ~i\cdot2^{n(I(U;Y_1|Z)-R_1-\e)}] \} \label{bins1} \\ 
C(j,m) &= \{V^n(l,m)~|~ l \in [(j-1)\cdot 2^{n(I(V;Y_2|Z)-R_2-\e)}+1,
~j\cdot2^{n(I(V;Y_2|Z)-R_2-\e)}] \}
\label{bins2}
\end{align}
where without loss of generality $2^{n(I(U;Y_1|Z)-R_1-\e)}$ and
$2^{n(I(V;Y_2|Z)-R_2-\e)}$ are considered to be integers, and
$[a,b]$ denotes the set of integers from $a$ to $b$. 
This imposes the following constraints on $R_1$ and $R_2$:
$R_1 \leq I(U;Y_1|Z)-\e$, and  $R_2 \leq I(V;Y_2|Z)-\e$.

\begin{figure}[h]
\centering \epsfig{file=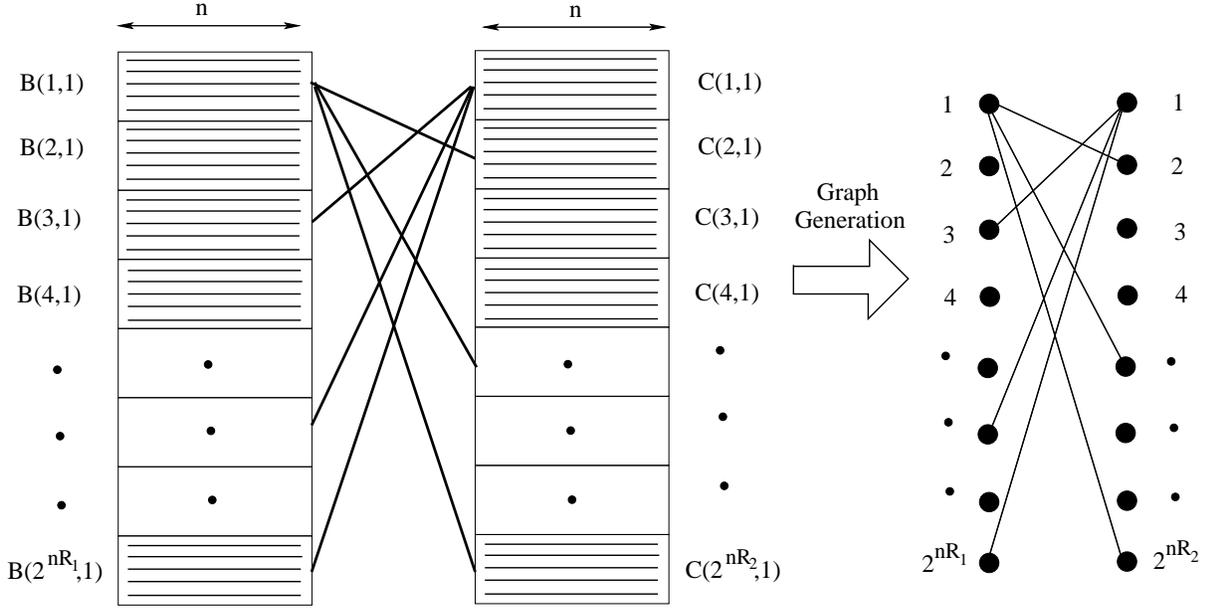, clip=,
width=.9\linewidth} \centering \caption{\small A subgraph
$\tl{\mb{G}}_1$ ($m=1$) with parameters $(2^{nR_1}, 2^{nR_2},
2^{nR_1'}, 2^{nR_2'}, \mu)$. }
\label{fig:bin-index-subgraph}
\end{figure}

\noindent \textbf{Graph generation}: As shown in Figure
\ref{fig:bin-index-subgraph}, for each $m$ in $\{1,2,\ldots,
2^{nR_0}\}$, a random graph 
$\tl{\mb{G}}_m$ can be associated with the bins as follows. (1)
$V_1(\tl{\mb{G}}_m)=\{(m-1)2^{nR_1}+1, \ldots, m2^{nR_1}\}$ and
$V_2(\tl{\mb{G}}_m)=\{(m-1)2^{nR_2}+1, \ldots, m2^{nR_2}\}$, (2)
$\forall (i,j) \in V_1(\tl{\mb{G}}_m) \times
V_2(\tl{\mb{G}}_m)$, $(i,j) \in E(\tl{\mb{G}}_m)$ if and only if there
exists in $B(i,m) \times C(j,m)$ at least one $\e$-strongly jointly
typical sequence pair 
that belongs to $A(U,V|Z^n(m))$, where 
$A(U,V|z^n)$ is the set of pairs of sequences
that are jointly $\e$-typical with the  sequence $z^n$, i.e., for $z^n
\in \Ae(Z)$, the set  
$A(U,V|z^n)$$ \triangleq \{(u^n,v^n)|(u^n,v^n,z^n)\in \Ae(U,V,Z)\}$.
Let $\mb{G}$ denote the random graph that is the union of 
$2^{nR_0}$ random graphs $\tl{\mb{G}}_m$ for $m
\in \{1,2,\ldots,2^{nR_0}\}$.

\vspace{10pt}

\noindent \textbf{Codebook generation}: A random channel codebook $\mb{C}$
can be generated from the graph $\mb{G}$ as follows. For every $m
\in \{1,2,\ldots,2^{nR_0}\}$,  and every $(i,j) \in
E(\tl{\mb{G}}_m)$, first find one pair of sequences $(U^n(k,m),V^n(l,m))
\in A(U,V|Z^n(m)) \cap (B(i,m) \times C(j,m))$. Then draw a random 
codeword $X^n(m,i,j)=f(m,i,j)$ uniformly from
$A(X|U^n(k),V^n(l),Z^n(m))$, where 
$A(X|u^n,v^n,z^n) \triangleq \{x^n| (x^n,u^n,v^n,z^n) \in \Ae (X,U,V,Z)\}$.
Thus the size of the
codebook is equal to the size of the edge set of the graph
$\Bbb{G}$, i.e., $|\mb{C}|=|E(\mb{G})|$.

\vspace{10pt} \noindent \textbf{Encoding error events due to the
degree condition}: Before we proceed to the encoding and decoding
procedure, we need to make sure that the generated codebooks satisfy
certain properties. If vertexes of the graph $\mb{G}$ do not satisfy
the degree conditions, the message pairs can not be transmitted with
arbitrarily small probability of error. Let 
\begin{equation}
A=I(U;Y_1|Z)+I(V;Y_2|Z)-I(U;V|Z)-2 \epsilon
\label{A},
\end{equation}
and choose $R'_1=A-R_2$ and $R'_2=A-R_1$. Since $0 \leq R'_i \leq R_i$
for $i=1,2$, we have the following conditions on $R_1$ and $R_2$:
\begin{equation}
\max\{R_1,R_2\} \leq A \leq R_1+R_2.
\label{AA}
\end{equation}

In summary,  the nonnegative tuple $(R_1,R_2,R'_1,R'_2)$ that we use
for the random coding  satisfies:
\beq
R_1 \leq I(U;Y_1|Z)-\e, \ \ \ R_1 \leq I(V;Y_2|Z)-\e, 
\label{constraint1}
\eeq
\beq
R_1+R'_2=R_2+R'_1= I(U;Y_1|Z)+I(V;Y_2|Z)-I(U;V|Z)-2 \e.
\label{constraint2}
\eeq

For a precise characterization of the error events, we need a function of 
$\epsilon$, and in turn certain properties of typical sets. For any 
triple $(U,V,Z)$ of finite-valued random variables, there
exists \cite{csiszar-korner} a continuous positive function
$\epsilon_1(\epsilon)$ (that depends on the triple)  
such that (a) $\epsilon_1(\epsilon) \rightarrow 0$ as $\epsilon
\rightarrow 0$ and (b) for all $\epsilon>0$ (sufficiently small), 
there exists an integer $N_0(\epsilon)>0$ such that $\forall
n>N_0(\epsilon)$ the following conditions hold simultaneously 
\begin{itemize}
\item  for all $z^n \in \Ae(Z)$,
\beq
2^{n(H(U|Z) -\epsilon_1)} \leq |A(U|z^n)| \leq
2^{n(H(U|Z)+\epsilon_1)}
\label{eqU}
\eeq
\beq
2^{n(H(V|Z) -\epsilon_1)} \leq |A(V|z^n)| \leq
2^{n(H(V|Z)+\epsilon_1)}
\label{eqV}
\eeq
\beq
2^{n(H(U,V|Z) -\epsilon_1)} \leq |A(U,V|z^n)| \leq
2^{n(H(U,V|Z)+\epsilon_1)}
\label{eqUV}
\eeq
\item  $\forall (u^n,z^n) \in \Ae(U,Z)$ 
\beq
2^{n(H(V|U,Z) -\epsilon_1)} \leq |A(V|u^n,z^n)| \leq
2^{n(H(V|U,Z)+\epsilon_1)}
\label{eqV|U}
\eeq
\item  $\forall (v^n,z^n) \in \Ae(V,Z)$ 
\beq
2^{n(H(U|V,Z) -\epsilon_1)} \leq |A(U|v^n,z^n)| \leq
2^{n(H(U|V,Z)+\epsilon_1)}.
\label{eqU|V}
\eeq
\end{itemize}

Coming back to the encoding error event, an error will be
declared if either one of the following events occur.
Let $\e'>3\e_1>0$.
\begin{itemize}
  \item $E_1$: $\exists i \in V_1(\mb{G})$ such that $\left|
  \frac{1}{n}\log{\rm deg}_{\mb{G},1}(i)-R_2' \right| > \e'$, 
  \item $E_2$: $\exists j \in V_2(\mb{G})$ such that $\left|
  \frac{1}{n}\log{\rm deg}_{\mb{G},2}(j)-R_1' \right| >   \e'$.
\end{itemize}

Now we show that the probability of these error events can be made
arbitrarily small under certain conditions. Let us define four events
as follows. 
\begin{itemize}
\item $E_{0,1}$ : $\exists i \in V_1(\mb{G})$ such that
${\rm deg}_{\mb{G},1}(i) <
2^{n(R'_2-\e')}$
\item $E_{0,2}$ : $\exists j \in V_2(\mb{G})$ such that ${\rm
deg}_{\mb{G},2}(j) <
2^{n(R'_1-\e')}$
\item $E^*_{0,1}$ : $\exists i \in V_1(\mb{G})$ such that ${\rm
deg}_{\mb{G},1}(i) >
2^{n(R'_2+\e')}$
\item $E^*_{0,2}$ : $\exists j \in V_2(\mb{G})$ such that ${\rm
deg}_{\mb{G},2}(j) >
2^{n(R'_1+\e')}$
\end{itemize}

Toward proving the required statements, we need a lemma given in
\cite{krithivasan} about certain properties of typical sets.
\begin{lem}\label{lem:bc-cc-0}
For any triple of finite-valued random variables $(U,V,Z)$, 
any $\epsilon>0$ (sufficiently small), any two positive
real numbers $R_1$ and $R_2$ such that $R_1+R_2>I(U;V|Z)$,  
and any  $z^n \in \Ae(Z)$ consider the following random experiment.
Generate two collections of sequences $\C_U(z^n)$ and $\C_V(z^n)$ 
of size $2^{nR_1}$ and $2^{nR_2}$ from $A(U|z^n)$ and
$A(V|z^n)$, respectively, with uniform distribution and with
replacement. Let $P_{\epsilon}(z^n,R_1,R_2)$ denote the probability that 
$|(\C_U \times \C_V) \cap A(U,V|z^n)|=0$. Then $\forall \epsilon>0$, 
$\forall R_1,R_2$ with $R_1+R_2>I(U;V|Z)$ and 
$\forall z^n \in \Ae(Z)$, 
\beq
\lim_{n \rightarrow \infty} -\frac{1}{n} \log P_{\epsilon}(z^n,R_1,R_2)=
\infty.
\eeq
\end{lem}

\noindent {\it Proof:} See Theorem 2.1 of \cite{krithivasan}.

The proofs of the next two lemmas use this result. 
\begin{lem}\label{lem:bc-cc-1}
  For any $\e>0$, and sufficiently large $n$:\\
\begin{equation}
 P\{E_{0,i}\} < \frac{\e}{12}, ~~\mbox{for $i=1,2$}
\end{equation}
\end{lem}

\noindent {\it Proof}: Refer to Appendix \ref{appendix:bc-lem-cc-1}.

\begin{lem}\label{lem:bc-cc-2}
  For any $\e>0$, and sufficiently large $n$:\\
\begin{equation}
 P\{E_{0,i}^*\} < \frac{\e}{12}, ~~\mbox{for $i=1,2$}
\end{equation}
\end{lem}

\noindent {\it Proof}: Refer to Appendix \ref{appendix:bc-lem-cc-2}.

Note that $E_1=E_{0,1} \cup E_{0,1}^*$ and $E_2=E_{0,2} \cup
E_{0,2}^*$. Thus, according to the Lemma \ref{lem:bc-cc-1} and  
\ref{lem:bc-cc-2}, it is easy to see that, for sufficiently large
$n$, $P(E_1) < \frac{\e}{6}$, and  $P(E_2) < \frac{\e}{6}$,
if $R_2'=I(U;Y_1|Z)-R_1+I(V;Y_2|Z)-I(U;V|Z)-2\e$ and
$R_1'=I(U;Y_1|Z)-R_2+I(V;Y_2|Z)-I(U;V|Z)-2\e$, respectively.
So, it is shown that with high probability we can obtain a nearly
semi-regular bipartite graph $\mb{G}$ composed of $2^{nR_0}$
disjoint subgraphs $\tl{\mb{G}}_m$ such that for each  $m$,  each vertex
in $V_1(\tl{\mb{G}}_m)$ has degree nearly equal to $2^{nR'_2}$
and each vertex in $V_2(\tl{\mb{G}}_m)$ has degree nearly equal to
$2^{nR'_1}$. The size of $V_1(\tl{\mb{G}}_m)$ is $2^{nR_1}$ and 
that of $V_2(\tl{\mb{G}}_m)$ is $2^{nR_2}$.

\vspace{10pt}\noindent \textbf{Choosing message correlation}: If
none of the above two error events $E_1$ and $E_2$ occurs, choose $G=\mb{G}$. 
Clearly  $G$ has parameters $(2^{nR_0}$,
$2^{nR_1}$, $2^{nR_2}$, $2^{nR_1'}$, $2^{nR_2'}$, $2^{n\e'})$.
If any of the
above two error events occurs, then pick any graph with parameters
$(2^{nR_0}$, $2^{nR_1}$, $2^{nR_2}$, $2^{nR_1'}$, $2^{nR_2'}$,
$2^{n\e'})$ and call it as $G$.
The distribution of $W_0$, $W_1$ and $W_2$ is chosen as follows.
$Pr\{(W_0,W_1,W_2)=(m,i,j)\}=\frac{1}{E(G)}$ if
$((m-1)2^{nR_1}+i,(m-1)2^{nR_2}+j)\in E(G)$ and
$Pr\{(W_0,W_1,W_2)=(m,i,j)\}=0$ else. For this  graph $G$, and the given
broadcast channel, using the above random codebook $\Bbb{C}$, we
construct an $(n,\tau)$-transmission system, where $\tau$ will be
specified in the sequel.

\vspace{10pt}\noindent \textbf{Encoding}: Sender transmits the
codeword $X^n(m,i,j)$ over the channel to deliver two pair of
messages $(m,i)$ and $(m,j)$ to receiver 1 and receiver 2,
respectively.

\vspace{10pt}\noindent \textbf{Decoding}: Both Receiver 1 and 
Receiver 2
first find the unique index $\hat{m}$ such that $Z^n(\hat{m})$ is
jointly typical with received sequence $Y_1^n$ and $Y_2^n$,
respectively. Then, Receiver 1 finds the unique index $\hat{k}$ such
that $U^n(\hat{k},\hat{m})$ is jointly typical with the received sequence
$Y_1^n$ and $Z^n(\hat{m})$, i.e., $(U^n(\hat{k},\hat{m}), Y_1^n,
Z^n(\hat{m})) \in \Ae(U,Y_1,Z)$. Similarly, Receiver 2 finds the
unique index $\hat{l}$ such that 
$(V^n(\hat{l},\hat{m}),
Y_2^n, Z^n(\hat{m})) \in \Ae(V,Y_2,Z)$. Then, each receiver finds
the decoded private messages $\hat{i}$ and $\hat{j}$ such that
$U^n(\hat{k},\hat{m}) \in 
B(\hat{i},\hat{m})$ and $V^n(\hat{l},\hat{m}) \in C(\hat{j},\hat{m})$,
respectively. Otherwise, an error will be declared.

\vspace{10pt}\noindent \textbf{Probability of error analysis}: So,
the probability of error $P(E)$ can be given by
\begin{align}
  P(E)&=P(E_1 \cup E_2)P(E|E_1 \cup E_2)+P(E \cap E_1^c \cap E_2^c)\\
       &\leq P(E_1 \cup E_2)+P(E \cap E_1^c \cap E_2^c)
\end{align}
The second probability in the above equation can be bounded as given
in the following lemma.

\begin{lem} \label{lem:bc-cc-3}
For any $\epsilon>0$, and sufficiently large $n$, \beq P(E \cap
E_1^c \cap E_2^c) < \frac{2 \epsilon}{3} \eeq provided
\begin{align}
  R_0 &\leq \min\{I(Z;Y_1),I(Z;Y_2) \}-\e. \\
\end{align}
\end{lem}
{\it Proof}: Refer to Appendix \ref{appendix:bc-lem-cc-3}. \\
Therefore, by applying the union bound we have 
$P(E) \leq  P(E_1)+P(E_2)+P(E \cap E_1^c \cap E_2^c) \leq \e$.

Since in every realization of random codebooks, we have chosen a
nearly semi-regular graph $G$ with parameters $(2^{nR_0}$,
$2^{nR_1}$, $2^{nR_2}$, $2^{nR_1'}$, $2^{nR_2'}$, $2^{n\e'})$, and
averaged over the ensemble of random codebooks, the average
probability of error is smaller than $\e$, there must exist a graph
$G$ with parameters $(2^{nR_0}$, $2^{nR_1}$, $2^{nR_2}$,
$2^{nR_1'}$, $2^{nR_2'}$, $2^{n\e'})$ and a codebook such that the
average probability of error is smaller than $\e$. This is true only
under the condition given by the statement of the theorem. Hence,
the proof of Theorem \ref{thm:rate-region-bc-cor} has been
completed. \hfill $\blacksquare$

\subsection{The Capacity Region of a Broadcast Channel with One
Deterministic Component with Correlated
Messages}\label{sec:one-deterministic-bc-cor}

In this section, we consider the
capacity region of broadcast channels with one deterministic
component with correlated
messages, $\mc{\tilde{C}}_{BC}$, when there is no common message,
i.e., when $R_0=0$. In this case, as expected, we can provide a
converse coding theorem.

\begin{thm}\label{thm:one-det-bc-cor}
For a discrete memoryless semi-deterministic broadcast channel where
$y_1=f_1(x)$ and the probability of $y_2$ given $x$ is $p(y_2|x)$,
$\mc{\tilde{R}}_{BC}^* = \mc{\tilde{C}}_{BC}$ where
\begin{align}
\mc{\tilde{R}}_{BC}^* = \bigcup_{p(v,x)}\{(R_1, R_2, R_1',
  R_2'):~~~~~& \notag\\ 
  R_1 &\leq H(Y_1) \label{eq:12}\\
  R_2 &\leq I(V;Y_2) \label{eq:13}\\
  R_1 + R_2'= R_1'+R_2 &\leq H(Y_1|V)+I(V;Y_2)\}\label{eq:14}
\end{align}
for some $p(v,x)$ on ${\mathcal V} \times
{\mathcal X}$, where $V$ is an  
auxiliary random variable with finite alphabet  
${\mathcal V}$ such
that $V \rightarrow X \rightarrow (Y_1,Y_2)$, 
$|{\mathcal V}| \leq |{\mathcal X}|+2$.
\end{thm}
{\it Proof:} See Appendix \ref{appendix:one-det-bc-cor}.

\section{Representation of Correlated Sources into Graphs}\label{sec:bc-sc}

In this section, we consider the problem of  representation of
correlated sources into graphs for transmission over  broadcast
channels, i.e., source 
coding module  in broadcast channels with correlated sources.
This problem can be interpreted as Gray-Wyner problem with correlated
messages. In in this setup, the  source encoder can access both sources to be
transmitted, but the source decoders can not collaborate with each
other.  We consider two  correlated sources $(\mc{S},\mc{T},p(s,t))$.

\subsection{Summary of Results}

\begin{figure}[h]
\centering \epsfig{file=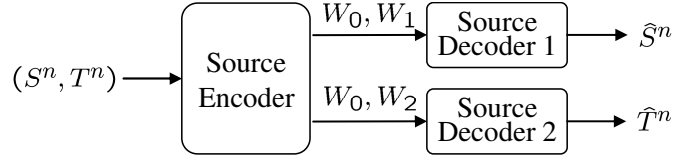, clip=, width=.5\linewidth}
\centering \caption{\small  Gray-Wyner problem with  correlated
messages} \label{fig:bc-sc-block}
\end{figure}

In this problem, the goal is to reliably represent two correlated
sources into a triple of messages $(W_0,W_1,W_2)$ which can be associated
with a nearly semi-regular bipartite graph $G$ with parameters
$(\Delta_0, \Delta_1, \Delta_2, \Delta_1',\Delta_2', \mu)$ as
defined in Definition \ref{def:bc-graph-5-parameters}. As shown in
Figure \ref{fig:bc-sc-block}, the output of source encoder is the triple
$W_0$, $W_1$ and $W_2$. We assume that two pairs of messages
$(W_0,W_1)$ and $(W_0,W_2)$ are sent to receiver 1 and receiver 2,
respectively, over the channel without error. From the received
message pairs, the two source decoders wish to reliably reconstruct the
original source sequences $S^n$ and $T^n$, respectively, without
communicating with each other.
For ease of exposition let is consider two simple functions
$I_1$ and $I_2$ both having domain as $\Bbb{Z} \times \Bbb{Z}$ and 
range as $\Bbb{Z}$ given by $I_1(a,b)=a$ and $I_2(a,b)=b$ for all
$(a,b) \in \Bbb{Z} \times \Bbb{Z}$. 

\begin{dfn}
An $(n,\tau)$-transmission system for a nearly semi-regular
bipartite graph $G$ with parameters $(\Delta_0$, $\Delta_1$,
$\Delta_2$, $\Delta_1'$, $\Delta_2'$, $\mu)$ and a pair of
correlated sources $({\mathcal S},{\mathcal T},p(s,t))$ 
is an ordered
tuple $(f,g_1,g_2)$, consisting of one encoding mapping $f$ and two
decoding mappings $g_1$ and $g_2$ where
\begin{itemize}
\item $f: \mc{S}^n \times \mc{T}^n \ra E(G)$,
\item $ g_1: V_1(G) \ra \mc{S}^n$, $g_2: V_2(G) \ra \mc{T}^n$,
 \item such that a performance measure given by the
probability of error satisfies:
\begin{align}
\tau&=Pr\{(g_1(I_1(f(S^n, T^n))),g_2(I_2(f(S^n, T^n)))) \!\neq\! (S^n,T^n) \}.
\end{align}
\end{itemize}
\end{dfn}

\noindent Now we define achievable rates for this problem as follows.
\begin{dfn}
  A tuple of rates $(R_0,R_1,R_2,R_1',R_2')$ is said to be
  \emph{achievable} for a pair of correlated sources
  $(\mc{S},\mc{T},p(s,t))$ (for 
  transmission over broadcast channels), if for all
$\e > 0$, and for all sufficiently large $n$,
  there exists a bipartite graph $G$ with parameters
  $(\Delta_0,\Delta_1, \Delta_2, \Delta_1', \Delta_2',\mu)$ and an
  associated
  $(n,\tau)$-transmission system as defined above satisfying:
  $\frac{1}{n}\log \Delta_i < R_i + \e$ for $i=0,1,2$,
  $\frac{1}{n}\log \Delta_j' < R_j' + \e$ for $j=1, 2$,
  $\frac{1}{n}\log \mu < \e$ 
  and the corresponding probability of error $\tau < \e$.
\end{dfn}

The goal is to find the achievable rate region $\mc{R}_{GW}$ which
is the set of all achievable tuple of rates $(R_0, R_1, R_2, R_1',
R_2')$. We have obtained an inner bound to the achievable rate region. It
is one of the main results of this paper and is given by the following theorem.

\begin{thm}\label{thm:bc-sc-main}
$\mc{R}_{GW}^* \subset \mc{R}_{GW}$ where
\begin{align}
\mc{R}_{GW}^* =\bigcup_{p(z|s,t)}\{(R_0, R_1, R_2, R_1', R_2')&: \notag\\
 R_0 &\geq I(S,T;Z), \label{eq:bc-sc-main1}\\
 R_1 &\geq H(S|Z),  \ \ \
 R_2 \geq H(T|Z), \label{eq:bc-sc-main3}\\
 R_1' &\geq H(S|T,Z), \ \ \
 R_2' \geq H(T|S,Z), \label{eq:bc-sc-main5}
\end{align}
where $Z$ is an auxiliary random variable with a finite alphabet 
$\mc{Z}$ such that $p(z,s,t)=p(s,t)p(z|s,t)$.

\end{thm}

\begin{rem}
Note that we can obtain $R_0+R_1+R_2'= R_0+R_1'+R_2 \geq H(S,T)$ by
combining the conditions in Theorem \ref{thm:bc-sc-main}. This
means that in every nearly semi-regular bipartite graph 
used to  represent the source, as expected,   
the total number of edges  must be greater than or equal to $2^{nH(S,T)}$.
\end{rem}

\subsection{Proof of Theorem
\ref{thm:bc-sc-main}}\label{sec:bc-sc-thm-proof}

In this section, we present a proof of Theorem
\ref{thm:bc-sc-main}. We use a random coding procedure and the notion of
strongly jointly typical sequences.
Let us consider a fixed finite set ${\mathcal Z}$, and a 
joint probability distribution $p(z,s,t)$ on
$\mc{Z} \times \mc{S} \times \mc{T}$.  Also fix $\e>0$, and real
numbers $R_0$, $R_1$, $R_2$, $R'_i=R_i-I(S;T|Z)$ for $i=1,2$.
Without loss of generality we assume that 
$R_i>I(S;T|Z)$ for $i=1,2$.

\vspace{10pt} \noindent \textbf{Random sequence generation}: First,
draw $2^{nR_0}$ sequences $Z^n(m)$, $m \in \{1, 2$, $\ldots,
2^{nR_0}\}$, of length $n$, independently from the strongly
$\e$-typical set $\Ae(Z)$. That is,
$P\{Z_m^n=z^n\}=\frac{1}{|\Ae(Z)|}$ if $z^n \in \Ae(Z)$, and
$P\{Z_m^n=z^n\}=0$ if $z^n \notin \Ae(Z)$. For every $m \in \{1,2,
\ldots, 2^{nR_0}\}$, draw $2^{nR_1}$  sequence  from $\Ae(S|Z^n(m))$ 
independently, equally likely and with replacement. Call this
collection $B(m)$. Denote the $i$th sequence as $S^n(i,m)$. 
By collecting  these $2^{nR_0}$ bins the first codebook $\Bbb{C}_1$ can be
obtained. In other words, $\Bbb{C}_1=\{B(1),B(2), \ldots,
B(2^{nR_0})\}$.

Similarly, for every  $m \in \{1,2,\ldots,2^{nR_0}\}$, 
construct  a bin $C(m)$ containing $2^{nR_2}$ sequences from 
$\Ae(T|Z^n(m))$. Let the $j$th sequence in $C(m)$ be denoted by $T^n(j,m)$.
The second codebook $\Bbb{C}_2$ can be
similarly generated, i.e., $\Bbb{C}_2=\{C(1),C(2), \ldots,
C(2^{nR_0})\}$.

\vspace{10pt} \noindent \textbf{Graph generation}: Now we generate a
bipartite graph $\Bbb{G}$ from the codebooks $\Bbb{C}_1$ and
$\Bbb{C}_2$ as follows:
\begin{itemize}
  \item $\Bbb{G}$ is composed of $2^{nR_0}$ disjoint subgraphs
  $\tl{\Bbb{G}}_m$ for $m \in \{1, 2, \ldots, 2^{nR_0}\}$,
  \item $V_i(\Bbb{G})\!=\!\cup_{m=1}^{2^{nR_0}}V_i(\tl{\Bbb{G}}_m)$
  for $i=1,2$, \ \ \ 
  $E(\Bbb{G})=\cup_{m=1}^{2^{nR_0}}E(\tl{\Bbb{G}}_m)$,
  \item $\forall m \!\in \!\{1, 2, \ldots, 2^{nR_0}\}$,
  $V_1(\tl{\Bbb{G}}_m)\!=\!\{(m-1)2^{nR_1}-1, \ldots, m2^{nR_1}\}$,  
$V_2(\tl{\Bbb{G}}_m)$=$\{(m-1)2^{nR_2}-1, \ldots, m2^{nR_2}\}$,
  \item $\forall m \!\in \!\{1, 2, \ldots, 2^{nR_0}\}$,
               $(i,j) \in E(\tl{\Bbb{G}}_m)$ if and only if 
$(S^n(i,m)$, $T^n(j,m)$, $Z^n(m))$ $\in
         \Ae(S,T,Z)$.
\end{itemize}

\vspace{10pt} \noindent \textbf{Encoding error events}: Before we
proceed further, let us make sure that the generated codebooks
satisfy certain properties. If the vertexes of $\Bbb{G}$ do not
satisfy certain degree requirements, we may not be able to reliably
represent the sources using this graph. 
For the triple $(S,T,Z)$, consider the function $\epsilon_1(\e)$ as 
defined in (\ref{eqU})-(\ref{eqU|V}). Let $\e'>3\e_1>0$. 
An encoding error will be declared if either one of the following
events occurs. \\
$E_1: \exists i \in V_1(\Bbb{G}) \mbox{ such } \mbox{ that }
| \frac{1}{n} \log {\rm deg}_{\Bbb{G},1}-R'_2 | > \e'$, \\
$E_2: \exists j \in V_2(\Bbb{G}) \mbox{ such } \mbox{ that }
| \frac{1}{n} \log {\rm deg}_{\Bbb{G},2}-R'_1 | > \e'$. 

Using the following lemma it can be shown that the probability of
these two events can be made arbitrarily small for large $n$. 

\begin{lem}\label{lem:bc-sc-1}
For any $\e>0$, and sufficiently large $n$,
we have 
$P(E_i)< \frac{\e}{8}, \ \ {\rm for } \ \ i=1,2.$
\end{lem}
{\it Proof:} The proofs of these two results are similar,
respectively, to those of Lemma
\ref{lem:bc-cc-1} and Lemma \ref{lem:bc-cc-2}. Hence for conciseness
we omit the proof.

So by Lemma \ref{lem:bc-sc-1}, $P(E_1)+P(E_2) < \frac{\e}{4}$. 
Hence we can obtain a bipartite graph
$\Bbb{G}$ where each vertex in $V_1(\Bbb{G})$ has degree nearly
equal to $ 2^{nR'_2}$ and each vertex in $V_2(\Bbb{G})$ has
degree nearly equal to $ 2^{nR'_1}$.

\vspace{10pt} \noindent \textbf{Choosing message correlation}: 
If none of the above error events $E_1$ and $E_2$ occurs, then choose
$G=\Bbb{G}$. If any of the above two error events occurs, then pick
any graph with parameters
$(2^{nR_0},2^{nR_1},2^{nR_2},2^{nR'_1},2^{nR'_2},2^{n\e'})$
and call it as $G$ and no guarantee will be given regarding the
probability of decoding error. 
For this graph $G$, and the given correlated sources $S$ and $T$, using
the above random codebooks $\Bbb{C}_1$ and $\Bbb{C}_2$, we construct
an $(n,\tau)$-transmission system, where $\tau$ will be specified in
the sequel.

\vspace{10pt} \noindent \textbf{Encoding}: If $(E_1\cup E_2)$ occurs,
then the encoder is some arbitrary mapping $\mc{S}^n \times \mc{T}^n
\rightarrow E(G)$. Otherwise, 
for a given $(S^n,T^n)
\!\in\! \Ae(S,T)$, find an index $m$, for $m \in \{1, 2,$ $\ldots,
2^{nR_0}\}$, such that $(S^n$, $T^n$, $Z^n(m))$ $\in \Ae(S,T,Z)$. If
there is no such index $m$, let $m$ be a random index chosen uniformly
from $\{1,2,\ldots,2^{nR_0}\}$. Also, find a pair of
indexes $(i,j)$ where $i$ (and $j$, respectively) is the index such
that $S^n(i,m)=S^n$ (and $T^n(j,m)=T^n$, respectively). If there is no
such sequence pair let $(i,j)$ be a random edge from the corresponding
sub-graph of $G$. The encoder sends $(m,i)$ and $(m,j)$ to
receiver 1 and receiver 2, respectively.

\vspace{10pt} \noindent \textbf{Decoding}: 
Given the received
index pair $(m,i)$, receiver 1 declares $\h{S}^n=S^n(i,m)$. Similarly,
given $(m,j)$, receiver 2 declares 
$\h{T}^n=T^n(j,m)$.

\vspace{10pt} \noindent \textbf{Probability of error analysis}: Let
$E$ denote the event $\{ (g_1(m,i),g_2(m,j)) \!\neq \!(S^n, T^n)\}$,
that the reconstruction vector pair is not equal to the source
vector pair. The probability of error $P(E)$ can be given by
\begin{align}
  P(E)&=P(E_1 \!\cup\! E_2)P(E|E_1 \! \cup \! E_2)\!+\!P(E \cap E_1^c
       \cap E_2^c)\\ 
       &\leq P(E_1 \!\cup\! E_2)+P(E \cap E_1^c \cap E_2^c).
\end{align}

The second probability in the above equation can be bounded as
given in the following lemma.

\begin{lem}\label{lem:bc-sc-3}
  For any $\e>0$, and sufficiently large $n$,
  \beq P(E \cap E_1^c \cap E_2^c) < \frac{3\e}{4} \eeq
  provided
  \beq
    R_0 > I(S,T;Z) + \e_2(\e), \ \ 
    R_1 > H(S|Z) + \e_1(\e), \ \ 
    R_2 > H(T|Z) + \e_1(\e),
  \eeq
  where $\e_2(\e) \ra 0$ as $\e \ra 0$ and $\e_2(\e)>0$.
\end{lem}
{\it Proof}: Refer to Appendix \ref{appendix:bc-lem-sc-3}.

Thus, $P(E)  \leq P(E_1)\!+\!P(E_2)\!+\! P(E \cap E_1^c \cap E_2^c) < \e$.
Therefore $P(E) < \e$ for sufficiently large $n$ and under the
conditions given by the theorem. In every realization of random
codebooks we have obtained a graph $G$ with the same set of parameters, and
averaged over this ensemble, we have made sure that the probability
of error is within the tolerance level of $\e$. Hence, the proof of
the direct coding theorem is completed. \hfill $\blacksquare$

\subsection{Outer bound}\label{sec:bc-sc-thm-proof}

In this section we provide a  partial converse for  Theorem
\ref{thm:bc-sc-main}.

\begin{thm}\label{thm:bc-sc-main_c}
$\mc{R}_{GW} \subset  \mc{R}_{GW}^{**}$ where
\begin{align}
\mc{R}_{GW}^{**} =\bigcup_{p(w|s,t)}\{(R_0, R_1, R_2, R_1', R_2')&: \notag\\
 R_0 &\geq I(S,T;Z), \label{eq:bc-sc-main1a}\\
 R_1 &\geq H(S|Z), \ \ \
 R_2 \geq H(T|Z), \label{eq:bc-sc-main3a}\\
R_1+R_2'&= R_1'+R_2 \geq H(S,T|Z)\label{eq:bc-sc-main6a}\},
\end{align}
where $Z$ is an auxiliary random variable such that $p(z,s,t)=p(s,t)p(z|s,t)$.
\end{thm}
{\it Proof:} Refer to Appendix \ref{appendix:bc-sc-main_c}.

\subsection{An interpretation of the rate region}\label{sec:different}

\begin{figure}[!h]
\begin{center}
\psfig{figure=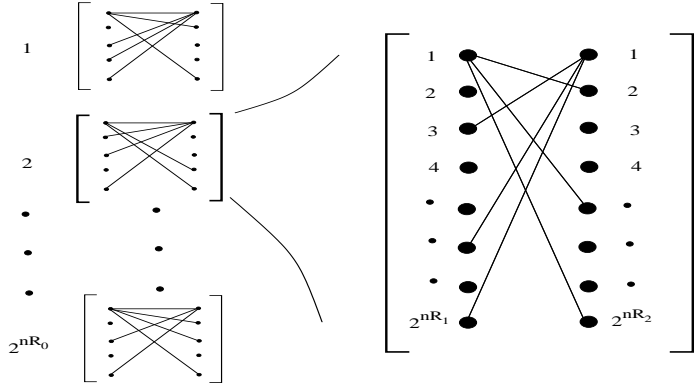,height=2.0in,width=3.6in}
\end{center}
\centering \caption{\small Graph $G$ with parameters $(2^{nR_0}$,
$2^{nR_1}$, $2^{nR_2}$, $2^{nR_1'}$, $2^{nR_2'}$, $2^{n\e'})$
composed of $2^{nR_0}$ subgraphs $\tl{G}_m$ with parameters 
($2^{nR_1}$, $2^{nR_2}$,
$2^{nR_1'}$, $2^{nR_2'}$, $2^{n\e'}$)}\label{fig:bc-message-graph}
\end{figure}

We have shown that for sufficiently large
block length $n$, a pair correlated sources $(S,T)$ can be
represented into  a nearly semi-regular graph $G$ with parameters
$(2^{nR_0}$, $2^{nR_1}$, $2^{nR_2}$, $2^{nR_1'}$, $2^{nR_2'}$,
$2^{n\e'}$) as shown in Figure \ref{fig:bc-message-graph}.
Note that many different graphs can be used to represent a pair
correlated sources without increasing the amount of ``global''
redundancy, i.e., the sum rate satisfies:
$R_0+R_1+R_2'=H(S,T)$. This can be seen from the fact that the total
number of edges in 
the graph satisfies:  $|E(G)|\approx 2^{n(I(S,T;Z)+H(S,T|Z))}=2^{nH(S,T)}$.
Further, the rate of the common message can vary from $0$ to $H(S,T)$.
As examples, let us consider some special cases as follows.
\begin{itemize}
  \item Case $A$: If $I(S,T;Z)=0$, then $R_0=0$, $R_1=H(S)$, $R_2=H(T)$,
  $R_1'=H(S|T)$, $R_2'=H(T|S)$. Roughly speaking, this
  corresponds to the typicality-graph of $(S,T)$.
  \item Case $B$: If $I(S,T;Z)=H(S,T)$, then $R_0=H(S,T)$, $R_1=0$, $R_2=0$,
  $R_1'=0$, $R_2'=0$.
\end{itemize}
In particular, consider the case when  $R_0=C(S;T)$, where $C(S;T)$
denotes the common information of  Wyner
\cite{wyner75}. 
In this case, $R_0=C(S;T)$, $R_1=H(S|Z)$, $R_2=H(T|Z)$,
  $R_1'=H(S|Z)$, $R_2'=H(T|Z)$ since $S$ and $T$ are conditionally
independent given $Z$. So, each subgraph $\tl{G}_m$ becomes
 nearly complete. This also means that the private messages for receiver 1
and 2 become nearly independent. Note that a subgraph $\tl{G}_m$ can
not be complete if $R_0 < C(S;T)$ because, by the definition,
$C(S;T)$ is the infimum of $I(S,T;Z)$ such that $S \ra Z \ra T$.

In summary, Case $A$ can be thought of as situated at one end of the
spectrum, and Case $B$ as situated on the other end of the spectrum.
For every value of $I(S,T;Z)$, we get an equally efficient
representation of the sources into a nearly semi-regular graph.
Here, an efficient representation means that the cardinality
of the edge set of a graph used for representation (of $n$ source
samples) is nearly equal to  $2^{nH(S,T)}$.

\section{Interpretation of the Example in Section
  \ref{sec:bc-ex}}\label{sec:bc-example} 

In this section, we revisit the example of Section \ref{sec:bc-ex}.
We analyze and
interpret this example from the perspective  of graphs and the
proposed coding scheme. This example deals with  transmission of
two correlated sources over a deterministic broadcast channel. This
example can be considered as a special case where the
typicality-graph of the sources and that of channel outputs can
match exactly. So, we can apply our coding theorem to this case.

Consider the pair of correlated sources of Section \ref{sec:bc-ex}:
$\mc{S}=\mc{T}=\{0, 1\}$ with distribution  $p(0,0)=p(0,1)=p(1,1)=1/3$, and the
Blackwell channel with ${\mathcal X}=\{1,2,3\}$, and ${\mathcal
  Y}_i=\{0,1\}$ for $i=1,2$ with conditional distribution 
$p(0,0|1)=p(0,1|2)=p(1,1|3)=1$.  Let $\alpha=h_2(1/3)$ and $\beta=\log 3
-h_2(1/3)$. It follows that $H(S)=H(T)=\alpha$
and $H(S|T)=H(T|S)=\beta$. Consider the rate tuple
$(0,\alpha,\alpha,\beta,\beta)$. Clearly this point belongs to
${\mathcal R}^*_{GW}$ as this tuple corresponds to the
typicality-graph of $({\mathcal S},{\mathcal T},p(s,t))$. 
Note that the above broadcast channel is a deterministic broadcast
channel. Using the arguments presented in Section
\ref{sec:one-deterministic-bc-cor}, it can be shown that for any
deterministic broadcast channel, for the special case when $R_0=0$,  
the capacity region $\tilde{{\mathcal C}}_{BC}$ is given by 

\beq
\bigcup_{p(x)} \{(R_1,R_2,R'_1,R'_2): \ \
R_i \leq H(Y_i) \ \ \mbox{for} \ \ i=1,2, \ \  
R_1+R'_2=R_2+R'_1 \leq H(Y_1,Y_2) \}.
\eeq

Choosing $X$ equally likely over $\{1,2,3\}$, it follows that 
the rate tuple $(\alpha,\alpha,\beta,\beta)$ belongs to 
$\tilde{{\mathcal C}}_{BC}$, which, of course,  implies that 
$(0,\alpha,\alpha,\beta,\beta)$ belongs to 
${\mathcal C}_{BC}$. In this case the distribution of the channel
output $(Y_1,Y_2)$ is the same as the distribution of the sources
$(S,T)$. Further, the graph corresponding to the tuple
$(0,\alpha,\alpha,\beta,\beta)$ in source coding is the same as the
graph corresponding to the tuple $(0,\alpha,\alpha,\beta,\beta)$ in channel
coding.  

In summary: (1) there is a
nonempty intersection between the achievable rate regions of the source
coding module and the channel coding module, and (2) the graph $G$
associated with the source coding module matches (is identical) with the
graph $G$ of channel coding module. 
Hence the given sources $(S,T)$
can be reliably sent over the given broadcast channel.

\section{Conclusion}\label{sec:bc-conclusion}
We have considered the problem of transmission of correlated sources over
broadcast channels. We have considered a graph-based modular architecture
involving two components: a channel coding component and a source
coding component.  The graphs are used to model the correlation between the
messages. Correlated sources 
are first mapped into such graphs, and the edges coming from these
graphs are reliably transmitted over a broadcast channel.

We have given a partial characterization of the set of all graphs
that can be used to represent a given pair of correlated sources,
and similarly given a partial characterization of the set of all
graphs such that edges coming those graphs are reliably transmitted
over a given broadcast channel.
We have also considered  special cases such as 
 deterministic broadcast channels and broadcast channels with one
 deterministic component, where converse results are provided. We 
have applied this analysis to the case of transmission of a triangular source
over a Blackwell channel as an example. The goal of
this work is to show that graphs may be used as discrete interface
in this modular approach to multiterminal communication problems.

\appendix

\section*{Appendix}

\section{Proof of Lemma \ref{lem:bc-cc-1}}\label{appendix:bc-lem-cc-1}
The event $E_{0,1}$ can be considered as
\begin{equation}
E_{0,1}=\bigcup_{i=1}^{2^{nR_1}} \bigcup_{m=1}^{2^{nR_0}} E_{0,1}(i,m)
\end{equation}
where  $E_{0,1}(i,m)$ is the event that 
${\rm deg}_{\mb{G},1}((m-1)2^{nR_1}+i) < 2^{n(R'_2-\e')}$.

Note that  each vertex in $V_1(\mb{G})$ is only connected with a
subset of $V_2(\tl{\mb{G}}_m)$ for some $m$, $m \in
\{1,2,\ldots,2^{nR_0}\}$.  
We define super-bins $\mf{B}(p,m)$ and $\mf{C}(q,m)$ for
$p\in [1,2^{n(R'_1-\epsilon')}]$ and $q\in[1,2^{n(R'_2-\epsilon')}]$, each of
which is a union of $2^{n(R_1+R_2-A+\epsilon')}$ consecutive bins $B(i,m)$ and
$C(j,m)$, respectively, i.e.,
\begin{align}
  \mf{B}(p,m)=\bigcup_{i=(p-1)2^{n(R_1+R_2-A+\epsilon')}+1}
^{p\,2^{n(R_1+R_2-A+\epsilon')}} B(i,m),\\
  \mf{C}(q,m)=\bigcup_{j=(q-1)2^{n(R_1+R_2-A+\epsilon')}+1}
^{q2^{n(R_1+R_2-A+\epsilon')}} C(j,m)
\end{align}
The size of each super-bin $\mf{B}(p,m)$ and
$\mf{C}(q,m)$ is $2^{n(I(U;Y_1|Z)-\epsilon-R'_1+\epsilon')}$ and
$2^{n(I(V;Y_2|Z)-\epsilon-R'_2+\epsilon')}$, respectively. Recall that 
$R_1+R'_2=R_2+R'_1=A$. 

Before we proceed further, let us observe that 
 $B(i,m)$, $\mf{B}(p,m)$ and $\C_1(m)$ are
collections of $2^{n(I(U;Y_1|Z)-\epsilon-R_1)}$,  
$2^{n(I(U;Y_1|Z)-\epsilon-R'_1+\epsilon')}$ and 
$2^{n(I(U;Y_1|Z)-\epsilon)}$, respectively, of random sequences.
Then, the event $E_{0,1}(i,m)$ can be expressed as
\begin{equation}
E_{0,1}(i,m) \subset \bigcup_{q=1}^{2^{n(R'_2-\e')}} E_{0,1}(i,q,m)
\end{equation}
where  $E_{0,1}(i,q,m)$ is the event that $(B(i,m) \times
\mf{C}(q,m)) \cap A(U,V|Z^n(m))$ is empty.  So, 
by using the union bound, the probability of this event
$P(E_{0,1}(i,m))$ can be bounded as:
\begin{align}
  P(E_{0,1}(i,m)) &\leq P \left( \bigcup_{q=1}^{2^{n(R'_2-\e')}}
  E_{0,1}(i,q,m) \right)\\ 
  &\leq \sum_{q=1}^{2^{n(R'_2-\e')}} P(E_{0,1}(i,q,m))\\
  &\leq \sum_{q=1}^{2^{n(R'_2-\e')}} \sum_{z^n \in \Ae(Z)}
  \frac{1}{|\Ae(Z)|} P(E_{0,1}(i,q,m)|Z^n(m)=z^n)
\end{align}
Therefore, for sufficiently large $n$, the probability of the event
$E_{0,1}$ can be bounded by applying the union bound:
\begin{align}
 P(E_{0,1}) &\leq \sum_{i=1}^{2^{nR_1}} \sum_{m=1}^{2^{nR_0}} P(E_{0,1}(i,m))\\
            &\leq \sum_{i=1}^{2^{nR_1}} \sum_{m=1}^{2^{nR_0}} 
 \sum_{q=1}^{2^{n(R'_2-\e')}} \sum_{z^n \in \Ae(Z)}
  \frac{1}{|\Ae(Z)|} P(E_{0,1}(i,q,m)|Z^n(m)=z^n) \\
              &\overset{(a)}\leq 2^{n(R_0+R_1+R'_2-\e')} 2^{-nM}\\
              &{\leq} \frac{\e}{12},
\end{align}
where (a) is from Lemma \ref{lem:bc-cc-0} because
\beq
\frac{1}{n} \log |B(i,m)||\mf{C}(q,m)|=I(U;V|Z)+\epsilon',
\eeq 
 and $M > 0$ is a
sufficiently large number satisfying $M > R_0+R_1+R'_2-\e'$.

In a similar way, we can also show that $P(E_{0,2})
<\frac{\e}{12}$ for sufficiently large $n$.
 \hfill $\blacksquare$

\section{Proof of Lemma \ref{lem:bc-cc-2}}\label{appendix:bc-lem-cc-2}

The event $E_{0,1}^*$ can be expressed as
\begin{equation}
E_{0,1}^*=\bigcup_{i=1}^{2^{nR_1}} \bigcup_{m=1}^{2^{nR_0}} E_{0,1}^*(i,m)
\end{equation}
where  $E_{0,1}^*(i,m)$ is the event that 
${\rm deg}_{\mb{G},1}((m-1)2^{nR_1}+i) > 2^{n(R'_2+\e')}$.

Let $\mc{D}(i,m)=\{j:  |A(U,V|Z^n(m))\cap (B(i,m) \times C(j,m))|
\neq 0 \}$. Then, 
\begin{align}
  |\mc{D}(i,m)|=\sum_{j=1}^{2^{nR_2}} \psi(j),
  \mbox{  where} ~~\psi(j)=\left\{%
\begin{array}{ll}
    1, & \hbox{if} \ \ |A(U,V|Z^n(m))\cap (B(i,m) \times C(j,m))| \neq 0, \\
    0, & \hbox{otherwise.} 
\end{array}%
\right.
\end{align}

In particular,
\begin{align}
  P\{\psi(j)=1\}&= P \{ |A(U,V|Z^n(m)) \cap
  (B(i,m) \times C(j,m))| \neq 0 \}\\
 &= \sum_{z^n \in \Ae(Z)} \frac{1}{|\Ae(Z)|} P \{ |A(U,V|Z^n(m)) \cap
  (B(i,m) \times C(j,m))| \neq 0 | Z^n(m)=z^n\}\\
   &\overset{(a)}{\leq} \sum_{z^n \in \Ae(Z)} \frac{1}{|\Ae(Z)|}
2^{n(I(U;Y_1|Z)-R_1-\e)}
  2^{n(I(V;Y_2|Z)-R_2-\e)} 2^{-n(I(U;V|Z)-3\e_1)}\\ 
  &= 2^{n(A-R_1-R_2+3\e_1)}
\end{align} 
where (a) is obtained by applying the union bound, and
from the property of strongly jointly $\e$-typical sequences
\cite{cover-thomas}: for all $z^n \in \Ae(Z)$, for a randomly and
independently chosen $U^n 
\in A(U|z^n)$ and $V^n \in A(V|z^n)$, for sufficiently large
$n$, the probability that
 $(U^n, V^n) \in A(U,V|z^n)$ is bounded by
\begin{equation}\label{eq:bc-cc-3}
2^{-n(I(U,V|Z)+3\e_1)} \leq P\{(U^n, V^n) \in A(U,V|z^n)\} \leq
2^{-n(I(U,V|Z)-3\e_1)}
\end{equation} 
where we have used (\ref{eqU}), (\ref{eqV}) and (\ref{eqUV}).

So, the expectation of $|\mc{D}(i,m)|$ can be bounded as follows.
\begin{align}
  E|\mc{D}(i,m)| &=\sum_{j=1}^{2^{nR_2}} P\{\psi(j)=1\}\\
             &\leq 2^{nR_2}2^{n(A-R_1-R_2+3\e_1)}\\
             &\leq 2^{n(R'_2+3\e_1)} \label{eq:bc-cc-5}
\end{align}

Now,
\begin{align}
  P\{E_{0,1}^*(i,m)\}&=P\{|\mc{D}(i,m)|>
                    \underbrace{2^{n(A-R_1+\e')}}_{\triangleq a}\}\\ 
                    &\overset{(b)}{<}e^{-at}E\{ e^{t|\mc{D}|} \}
\end{align}
for any $t>0$ where $(b)$ follows from the Chernoff bound
\cite{cover-thomas}.

Now we calculate an upper bound for $E\{ e^{t|\mc{D}|}\}$. Consider an
arbitrary sequence $z^n \in \Ae(Z)$.   Let $u^n[l]$ be the
$l$-th sequence (using some ordering) in $A(U|z^n)$.
Recalling the definition given in (\ref{bins1}) and (\ref{bins2}),
let us denote $|B(i,m)|=\beta$, and 
$i_k=(i-1)\beta+k$ for $k \in \{1,2,\ldots,\beta\}$ for notational
simplicity. Now consider the following sequence of arguments:
\begin{align}
  E\{ e^{t|\mc{D}(i,m)|}|Z^n(m)=z^n \} 
  &= E \bigg\{ \prod_{j=1}^{2^{nR_2}} e^{t\psi(j)} \bigg| Z^n(m)=z^n 
 \bigg\}\\
  &= \sum_{l_1=1}^{|A(U|z^n)|}\!\!\!\!p\{U^n(i_1,m)=u^n[l_1]\} \!\!
  \sum_{l_2=1}^{|A(U|z^n)|}
  \!\!\!\!p\{U^n(i_2,m)=u^n[l_2]\}\ldots
  \!\!\sum_{l_{\beta}=1}^{|A(U|z^n)|}\!\!\!\!p\{U^n(i_\beta,m)=u^n[l_\beta]\}
  \notag\\ 
  &~~~~~~~~~~~E \bigg\{ \prod_{j=1}^{2^{nR_2}} e^{t\psi(j)}\bigg|
  ~\mbox{$U^n(i_\theta,m)=u^n[l_\theta]$, 
  for $\theta=1,2,\ldots,\beta$ }, \mbox{and }  Z^n(m)=z^n \bigg\}\\
  &= \sum_{l_1=1}^{|A(U|z^n)|}\sum_{l_2=1}^{|A(U|z^n)|}\ldots
  \sum_{l_{\beta}=1}^{|A(U|z^n)|}\frac{1}{|A(U|z^n)|^{\beta}} 
  \notag\\
  &~~~~~~~~~~~E \bigg\{ \prod_{j=1}^{2^{nR_2}}
  e^{t\psi(j)}\bigg|~\mbox{$U^n(i_\theta,m)=u^n[l_\theta]$, 
  for $\theta=1,2,\ldots,\beta$ }, \mbox{ and } Z^n(m)=z^n
  \bigg\}\\
  &\overset{(c)}{=}
  \sum_{l_1=1}^{|A(U|z^n)|}\sum_{l_2=1}^{|A(U|z^n)|}\ldots
  \sum_{l_{\beta}=1}^{|A(U|z^n)|}\frac{1}{|A(U|z^n)|^{\beta}} 
  \notag\\
  &~~~~~~~~~~~\prod_{j=1}^{2^{nR_2}} E \bigg\{
  e^{t\psi(j)}\bigg|~~\mbox{$U^n(i_\theta,m)=u^n[l_\theta]$, 
  for $\theta=1,2,\ldots,\beta$ }, \mbox{ and } Z^n(m)=z^n
  \bigg\}
\end{align}
where $(c)$ is from the fact that $\psi(j)$'s are independent when
the outcomes of $U^n(i_1,m),U^n(i_2,m),\ldots,U^n(i_\beta,m)$ and
$Z^n(m)$ are fixed. Let
us denote 
\beq
p_j=P\{\psi(j)=1 \big|~\mbox{$U^n(i_\theta,m)=u^n[l_\theta]$,
  for $\theta=1,2,\ldots,\beta$ }, \mbox{ and } Z^n(m)=z^n \}.
\label{pj}
\eeq

Then,
\begin{align}
  E \{e^{t\psi(j)} \big|~\mbox{$U^n(i_\theta,m)=u^n[l_\theta]$,
  for $\theta=1,2,\ldots,\beta$ }, \mbox{ and } Z^n(m)=z^n  \}&=e^t
  p_j+1\cdot (1-p_j)\\   
 &=1-p_j(1-e^t)\\
                    &\leq e^{-p_j(1-e^t)}~~~ \mbox{since $1-x \leq e^{-x}$} .
\end{align}
So,
\begin{align}
&\prod_{j=1}^{2^{nR_2}} E
\{e^{t\psi(j)}|~\mbox{$U^n(i_\theta,m)=u^n[l_\theta]$,
  for $\theta=1,2,\ldots,\beta$ }, \mbox{ and } Z^n(m)=z^n  \}\\
  &\leq \prod_{j=1}^{2^{nR_2}} e^{p_j(e^t-1)}\\
&={\rm exp} \bigg\{ (e^t-1) \sum_{j=1}^{2^{nR_2}} p_j\bigg\}\\
&={\rm exp} \bigg\{ (e^t-1) E\left\{|\mc{D}(i,m)| \big|
~\mbox{$U^n(i_\theta,m)=u^n[l_\theta]$,
  for $\theta=1,2,\ldots,\beta$ },  \mbox{ and } Z^n(m)=z^n \right\} \bigg\}
\end{align}

Then,
\begin{align}
 E\{ e^{t|\mc{D}(i,m)|} | Z^n(m)=z^n \}
 &\leq
\sum_{l_1=1}^{|A(U|z^n)|}\!\!\!\!\ldots\!\!\!\!
\sum_{l_{\beta}=1}^{|A(U|z^n)|}\!\!\!\!\frac{1}{|A(U|z^n)|^{\beta}}
 \notag\\ 
  &~~~~~~~~{\rm exp} \bigg\{ (e^t-1) E\left\{|\mc{D}(i,m)| \big|
\mbox{$U^n(i_\theta,m)=u^n[l_\theta]$,
  for $\theta=1,2,\ldots,\beta$ }, \mbox{ and } Z^n(m)=z^n  \right\}
 \!\! \bigg\}\\ 
  &\overset{(d)}{\leq} {\rm exp} \bigg\{ (e^t-1) 2^{n(R'_2+3\e_1)}  \bigg\}
\end{align}
where $(d)$ is obtained because $E\left\{|\mc{D}(i,m)| \big|
~\mbox{$U^n(i_\theta,m)=u^n[l_\theta]$,
  for $\theta=1,2,\ldots,\beta$ }, \mbox{ and } Z^n(m)=z^n  \right\}$
is bounded by $2^{n(R'_2+3 \e_1)}$ (the same as the bound on the
unconditional expectation as in inequality 
(\ref{eq:bc-cc-5}))  regardless of the particular sequence collection
$u^n[l_1],u^n[l_2],\ldots,u^n[l_\beta]$ and $z^n$. To see this crucial step 
clearly let us rewrite  $p_j$ (see (\ref{pj})), using the union bound
on the outcome of the bin $C(j,m)$, as follows: 
\beq
p_j \leq  2^{n(I(V;Y_2|Z)-R_2-\epsilon)} 
  P \left[  \bigcup_{\theta=1}^{\beta} F_{\theta}  \right],
\eeq
where $F_{\theta}$ is the event that a sequence drawn from $A(V|z^n)$ 
belongs to $A(V|u^n[l_{\theta}],z^n)$. Since all sequences in
$B(i,m)$, given that $Z^n(m)=z^n$, 
are drawn from $A(U|z^n)$, we have using (\ref{eqV}) and (\ref{eqV|U})
\beq
p_j \leq  2^{n(I(V;Y_2|Z)-R_2-\epsilon)}  2^{n(I(U;Y_1|Z)-R_1-\epsilon)} 
2^{-n(I(U;V|Z)-3 \epsilon_1)}= 2^{n(A-R_1-R_2+3\e_1)}.
\eeq

Therefore, for $t>0$,
\begin{align}
  P\{E_{0,1}^*(i,m)\} &\leq e^{-at} {\rm exp} \bigg\{ (e^t-1)
    2^{n(R'_2+3\e_1)}  \bigg\}\\ 
    &={\rm exp} \bigg\{-at+(e^t-1)\underbrace{2^{n(R'_2+3\e_1)}}_{\triangleq
    ~b}\bigg\}
\end{align}

To get a tighter upper bound, let us denote $f(t)=-at+b(e^t-1)$,
for $t>0$. Then, $f'(t)=-a+be^t$ and $f''(t)=be^t
> 0$. So, $f(t)$ has the minimum value when $t=\ln
\left(\frac{a}{b} \right)$.

Thus, $P\{E_{0,1}^*(i,m)\}$ is bounded as
\begin{align}
  P\{E_{0,1}^*(i,m)\}& \leq {\rm exp} \bigg\{-a\ln \left( \frac{a}{b}
  \right)+a-b \bigg\} 
\end{align}
where $a=2^{n(R'_2+\e')}$ and $b=2^{n(R'_2+3\e_1)}$.

So,
\begin{align}
  P\{E_{0,1}^*(i,m)\} &\leq {\rm exp} \left\{-2^{n(R'_2+3\e_1)}
  [2^{n(\e'-3\e_1)}\ln(2^{n(\e'-3\e_1)})-2^{n(\e'-3\e_1)}+1 ]
  \right\}\\*
  &= {\rm exp} \big\{-2^{n(R'_2+3\e_1)}\eta \big\} \\
  &\leq  {\rm exp} \big\{-2^{n(R'_2+3\e_1)} \big\}
\end{align}
where
$\eta=2^{n(\e'-3\e_1)}\ln(2^{n(\e'-3\e_1)})-2^{n(\e'-3\e_1)}+1$. Note
that  $\eta >0$ since (i) $\e'>3\e_1$ and (ii) for $x>1$, $g(x)=x\ln(x)-x+1$ is
increasing function of $x$ and $g(x)>0$. Further $g(e)=1$. 

Therefore, for sufficiently large $n$, by applying the union bound,
\begin{align}
P\{E_{0,1}^*\} &=P\Big\{ \bigcup_{i=1}^{2^{nR_1}} \bigcup_{m=1}^{2^{nR_0}} 
E_{0,1}^*(i,m) \Big\}\\
               &\leq \sum_{i=1}^{2^{nR_1}}\sum_{m=1}^{2^{nR_0}} 
 P\{E_{0,1}^*(i)\}\\
               &\leq 2^{n(R_0+R_1)}{\rm exp} \big\{-2^{n(R'_2+3\e_1)}
 \big\} \\ 
               &={\rm exp} \big\{n(R_0+R_1)\ln2-2^{n(R'_2+3\e_1)}
  \big\} \\ 
               &\overset{(e)}{<} \frac{\e}{12}
\end{align}
where (e) is from the fact that
$n(R_0+R_1)\ln2$ is linearly increasing but $2^{n(R'_2+3\e_1)}$ is
exponentially increasing as $n$ increases.

In a similar way, we can also show that
$P\{E_{0,2}^*\}<\frac{\e}{12}$ for sufficiently large $n$.
 \hfill $\blacksquare$

\section{Proof of Lemma \ref{lem:bc-cc-3}}\label{appendix:bc-lem-cc-3}
Now let us calculate the probability $P(E \cap E_1^c \cap E_2^c)$.
Without loss of generality, let us assume that the outcome of the 
message triple is given by: $(W_0,W_1,W_2)=(m,i,j)$. Let 
$(U^n(k_i,m),V^n(l_j,m))$ be the pair of sequences that are jointly 
typical in the bin $B(i,m)$ and $C(j,m)$.
Consider the following error events.
\begin{itemize}
  \item[$E_3$:] $(Z^n(m),U^n(k_i,m),V^n(l_j,m),X^n, Y_1^n, Y_2^n) \notin
  \Ae(Z,U,V,X,Y_1,Y_2)$ 
  \item[$E_4$:] Decoding step fails at receiver 1, i.e., $ \exists
  ~\hat{m} \neq m$ such that $(Z^n(\hat{m}), Y_1^n) \in 
  \Ae(Z,Y_1)$ or
  $ \exists ~\hat{k} \neq k_i$ such that $(U^n(\hat{k},m), Y_1^n, Z^n(m))
  \in A_{\e}^{(n)}(U,Y_1,Z)$ 
  \item[$E_5$:] Decoding step fails at receiver 2, i.e., $ \exists
  ~\hat{m} \neq m$ such that $(Z^n(\hat{m}), Y_2^n) \in 
  \Ae(Z,Y_2)$ or
  $ \exists ~\hat{l} \neq l_j$ such that $(V^n(\hat{l},m), Y_2^n, Z^n(m))
  \in A_{\e}^{(n)}(V,Y_2,Z)$ 
\end{itemize}

It is easy to see that, for sufficiently large $n$, 
$P(E_3) \leq \frac{2\e}{9}$, using  the Markov lemma
\cite{berger,cover-thomas};
$P(E_4) \leq \frac{2\e}{9}$, if $R_0 \leq I(Z;Y_1)-\e$ and  
$P(E_5) \leq \frac{2\e}{9}$, if $R_0 \leq I(Z;Y_2)-\e$.
Recall that for all $m \in \{1,2,\ldots,2^{nR_0}\}$, 
$|\C_1(m)|=2^{n(I(U;Y_1|Z)-\e)}$ and 
$|\C_2(m)|=2^{n(I(V;Y_2|Z)-\e)}$.

So,
\begin{align}
  P(E \cap E_1^c \cap E_2^c) &= P(\cup_{i=3}^5 E_i \cap E_1^c \cap
  E_2^c)\\
  &\leq \sum_{i=3}^5 P\left(E_i \cap E_1^c \cap E_2^c \right) <
  \frac{2\e}{3}
\end{align}
Hence, the proof of Lemma \ref{lem:bc-cc-3} has been completed.
\hfill $\blacksquare$

\section{Proof of Theorem \ref{thm:one-det-bc-cor}}
\label{appendix:one-det-bc-cor}

Here we provide the converse part, as the direct part follows from 
Theorem \ref{thm:rate-region-bc-cor}. The proof is very similar to 
Marton's proof \cite{marton79} of the
outer bound of the capacity region of semi-deterministic broadcast
channels. Consider any sequence (indexed by $n$) of $(n,\tau(n))$ 
transmission systems for a sequence of bipartite graphs $G_n$,
respectively, with parameters
$(1,2^{nR_1},2^{nR_2},2^{nR'_1},2^{nR'_2},\mu(n))$ and the
broadcast channel with one deterministic component such that 
$\tau(n) \rightarrow 0$, and $\frac{1}{n} \log \mu(n) \rightarrow 0$
as $n \rightarrow \infty$. 

By Fano's Inequality,
\begin{align}
  H(W_1,W_2|Y_1^n,Y_2^n) ~&\leq ~ \tau(n) \cdot n(R_1+R_2) +
  H(\tau(n)) ~\leq~n\e_n \\ 
  H(W_1|Y_1^n) ~&\leq ~ \tau(n) \cdot n R_1 + H(\tau(n)) ~\leq~n\e_n \\
  H(W_2|Y_2^n) ~&\leq ~ \tau(n) \cdot n R_2 + H(\tau(n)) ~\leq~n\e_n
\end{align} where $\e_n \rightarrow 0$ as $\tau(n) \rightarrow 0 $.
We can bound the rate $R_1$ as
\begin{align}
  n R_1 &= H(W_1) = I(W_1;Y_1^n)+H(W_1|Y_1^n)\\
        &\leq I(W_1;Y_1^n)+ n\e_n 
        \leq H(Y_1^n) + n \e_n \\
        &\leq \sum_{i=1}^{n} H(Y_{1i}) + n\e_n\\
        \Rightarrow R_1 &\leq \frac{1}{n}\sum_{i=1}^{n} H(Y_{1i}) +
        \e_n. \label{eq:15} 
\end{align}
$R_2$ can be bounded as
\begin{align}
  nR_2 &= H(W_2) = I(W_2;Y_2^n)+H(W_2|Y_2^n)\\
        &\leq I(W_2;Y_2^n)+ n\e_n 
        =\sum_{i=1}^{n} I(W_2;Y_{2i}|Y_{2}^{(i-1)}) + n\e_n\\
	 &\leq \sum_{i=1}^{n}
        [H(Y_{2i})-H(Y_{2i}|W_2,Y_2^{i-1},Y_{1(i+1)}^n)] + n\e_n\\ 
        &=\sum_{i=1}^{n} I(V_i;Y_{2i}) + n\e_n, ~~\text{by defining
        $V_i=(W_2,Y_2^{i-1},Y_{1(i+1)}^n)$}\\ 
\Rightarrow R_2 &\leq \frac{1}{n}\sum_{i=1}^{n} I(V_i;Y_{2i}) +
\e_n. \label{eq:16}
\end{align}

Using the fact that (a) the channel is discrete memoryless and is used
without feedback, and (b) $Y_1$ is a deterministic function of the 
channel input, it can be easily shown that $V_i \rightarrow X_i
\rightarrow (Y_{1i},Y_{2i})$. 
Before we bound the sum of rates $R_1+R_2'$, let us recall the
following identity from \cite{gel80}: for any two sequence of 
random variables $A^n$ and $B^n$, 
\beq
\sum_{i=1}^n I(A^{(i-1)};B_i|B_{(i+1)}^n)=
\sum_{i=1}^n I(B_{(i+1)}^n;A_i|A^{(i-1)}).
\label{specialequality}
\eeq

The number of all possible pairs of messages $(W_1,W_2)$ which have
non-zero probability is lower
bounded by $2^{n(R_1+R_2')}\mu^{-1}=2^{n(R_1'+R_2)}\mu^{-1}$.
So, we can bound the sum rate $n(R_1+R_2')$ as
\begin{align}
  n(R_1+R_2') &\leq H(W_1,W_2) + \log \mu\\
        &=H(W_2)+H(W_1|W_2) + \log \mu\\
        &=I(W_2;Y_2^n)+\underbrace{H(W_2|Y_2^n)}_{\leq n\e_n}
        +I(W_1;Y_1^n|W_2)+\underbrace{H(W_1|Y_1^n,W_2)}_{\leq
        H(W_1|Y_1^n) \leq ~n\e^n} + \log \mu\\ 
        &\leq I(W_1;Y_1^n|W_2)+I(W_2;Y_2^n)+2n\e_n + \log \mu\\
        &= H(Y_1^n)-I(W_2;Y_1^n)+I(W_2;Y_2^n)+2n\e_n + \log \mu\\
        &\overset{(a)}{\leq} H(Y_1^n) -\left[ \sum_{i=1}^n (I(W_2,
        Y_{1(i+1)}^n; Y_{1i}) - I(Y_{1(i+1)}^n;Y_{1i})) \right] +
        \sum_{i=1}^n I(W_2,Y_2^{(i-1)}; Y_{2i})+2n\e_n + \log \mu\\
        &\overset{(b)}= \sum_{i=1}^n H(Y_{1i})+ \sum_{i=1}^n \left[
        I(V_i;Y_{2i})-I(V_i;Y_{1i}) \right] + 2n\e_n + \log \mu\\
        &= \sum_{i=1}^n \left[H(Y_{1i}|V_i)+I(V_i;Y_{2i}) \right] +
        2n\e_n + \log \mu\\ 
       \Rightarrow R_1 + R_2' &\leq \frac{1}{n}\sum_{i=1}^{n}
(H(Y_{1i}|V_i)+I(V_i;Y_{2i})) + 2\e_n + \frac{1}{n}\log \mu \label{eq:17}
\end{align}
where\\
$(a)$ follows from the fact that $I(W_2;Y_{2i}|Y_{2}^{(i-1)}) \leq 
I(W_2,Y_{2}^{(i-1)};Y_{2i})$. \\
$(b)$ from the fact that $H(Y_1^n)+\sum_{i=1}^n I(Y_{1(i+1)}^n;Y_{1i})=
\sum_{i=1}^n H(Y_{1i})$, and (\ref{specialequality}).

To complete the proof, let us define a new random variable $Q$
uniformly distributed over the set $Q=\{1,2,\ldots,n\}$ with
probability $\frac{1}{n}$ as shown in \cite{cover-thomas,gelfand80}. By
defining $V=V_Q$, $Y_1=Y_{1Q}$, $Y_2=Y_{2Q}$, it can be shown that
the inequalities (\ref{eq:12}), (\ref{eq:13}), and (\ref{eq:14})
are equivalent to (\ref{eq:15}), (\ref{eq:16}), and (\ref{eq:17}),
respectively. \hfill $\blacksquare$

\section{Proof of Lemma \ref{lem:bc-sc-3}}\label{appendix:bc-lem-sc-3}
Let us calculate the probability $P(E \cap E_1^c \cap E_2^c)$. If
previous error events $E_1$ or $E_2$ do not occur, we define other
error events as follows.
\begin{itemize}
  \item[$E_3$]: $(S^n,T^n) \notin \Ae$,
  \item[$E_4$]: $\nexists m \in \{1,2,\ldots,2^{nR_0}\}$ such that
  $(S^n,T^n,Z^n(m))\in \Ae(S,T,Z)$,
  \item[$E_5$]: $\nexists i \in \{1,2,\ldots,2^{nR_1}\}$ such that
    $S^n(i,m)=S^n$ and 
$S^n(i,m) \in B(m)$,
  \item[$E_6$]: $\nexists j \in \{1,2,\ldots,2^{nR_2}\}$ such that
    $T^n(j,m)=T^n$ and 
$T^n(j,m) \in C(m)$,
\end{itemize}
Then,
\begin{align}
  P(E \cap E_1^c \cap E_2^c)&=P\left( \cup_{i=3}^6 E_i \cap (E_1^c \cap
                                   E_2^c) \right)\\ 
     &\leq  P\left[E_3 \cap (E_1^c \cap E_2^c) \right] +
\sum_{i=4}^6 P\left(E_i \cap E^c_3 \cap  (E_1^c \cap E_2^c) \right) 
\end{align}

By the property of jointly typical sequences \cite{cover-thomas},
$P[E_3 \cap (E_1^c \cap E_2^c)]< \frac{\e}{8}$ for sufficiently large
$n$. Using arguments of Chapter 13 of \cite{cover-thomas}, it can be
shown that  for sufficiently large $n$, if $ R_0 > I(S,T;Z) + \e_2(\e)$ where
$\e_2(\e) \ra 0$ as $\e \ra 0$, then 
$P[E_4 \cap E_3^c \cap (E_1^c \cap E_2^c)] \leq \frac{\e}{8}$. 
 See Chapter 13 of \cite{cover-thomas}
for a characterization of $\e_2$. Now
\begin{align}
  P[E_5 \cap E_3^c \cap E_4^c \cap (E_1^c \cap E^c_2)]
  &\overset{(a)}{=}P[\forall S^n(i,m) \in B(m), S^n(i,m) \neq S^n | S^n
  \in A(S|Z^n(m))]\\ 
         &\overset{(b)}{\leq} \left[1-2^{-n(H(S|Z)+\e_1(\e))}
  \right]^{2^{nR_1}}\\ 
         &\overset{(c)}{\leq} {\rm exp}\{-2^{n(R_1-H(S|Z)-\e_1(\e))}\}\\
         &< \frac{\e}{8}
\end{align}
for sufficiently large $n$, if $R_1>H(S|Z)+\e_1(\e)$ where\\
(a) is from the fact that $(S^n,T^n,Z^n(m))\in \Ae(S,T,Z)$ implies
$S^n \in A(S|Z^n(m))$,\\
(b) is obtained using the fact that each $S^n(i,m)$ has the same chance of
equaling $S^n$ and independently chosen, and 
the fact that  $|A(S|Z^n(m))|\leq 2^{n(H(S|Z)+\e_1(\e))}$,\\
(c) follows from $(1-x)^n \leq e^{-xn}$ for $0 \leq x \leq 1$ and
$n>0$.

So,
\begin{align}
  P[E_5 \cap E_3^c \cap (E_1^c\cap E^c_2)]&=P[E_5 \cap E_3^c \cap E_4
                                   \cap (E_1^c \cap E^c_2)]
                                   + P[E_5 \cap E_3^c \cap E_4^c \cap
                                   (E_1^c \cap E^c_2)]\\ 
                                   &\leq P[E_4 \cap E_3^c \cap (E_1^c
                                   \cap E^c_2)]
                                   + P[E_5 \cap E_3^c \cap E_4^c \cap 
                                   (E_1^c \cap E_2^c)]\\
                                   &< \frac{\e}{4}
\end{align}

Similarly, it can be shown that $P[E_6 \cap E_3^c \cap (E_1^c \cap E^c_2)] <
\frac{\e}{4}$ if $R_2>H(T|Z)+\e_1(\e)$ for sufficiently large $n$.
Hence $P[E \cap(E_1^c \cap E_2^c)] \leq \frac{3 \e}{4}$.
\hfill $\blacksquare$

\section{Proof of Theorem \ref{thm:bc-sc-main_c}}
\label{appendix:bc-sc-main_c}

Let $f(n),g_1(n),g_2(n)$ be a fixed sequence of encoder and
decoders. Also, let $f(S^n,T^n)=(W_0,W_1,W_2)$. 
We can bound the rate $R_0$ as
\begin{align}
  nR_0 &\geq H(W_0)\geq I(S^n,T^n;W_0)\\
               &=\sum_{i=1}^n [H(S_i,T_i|S^{i-1},T^{i-1})-
               H(S_i,T_i|\underbrace{S^{i-1},T^{i-1},W_0}_{\triangleq
               Z_i})]\\ 
              &= \sum_{i=1}^n [H(S_i,T_i)- H(S_i,T_i|Z_i)]
              = \sum_{i=1}^n I(S_i,T_i;Z_i) \label{eq:bc-sc-12}
\end{align}
Using the constraints on the degrees of the vertexes in the graph
associated with messages, we can also have
\begin{align}
n(R_2+R'_1)=  n(R_1+R_2') &\overset{(a)}{\geq} H(W_1,W_2|W_0) - \log \mu\\
              &= I(S^n,T^n;W_1,W_2|W_0)+H(W_1,W_2|S^n,T^n,W_0) - \log \mu\\
              &\overset{(b)}{=} I(S^n,T^n;W_1,W_2|W_0) - \log \mu\\
              &=H(S^n,T^n|W_0)-H(S^n,T^n|W_1,W_2,W_0) - \log \mu\\
              &\overset{(c)}{\geq} H(S^n,T^n|W_0)-n\e_n - \log \mu \\
              &= \sum_{i=1}^n
              H(S_i,T_i|\underbrace{S^{i-1},T^{i-1},W_0}_{=Z_i})-n\e_n
              - \log 
              \mu,\\
              &= \sum_{i=1}^n H(S_i,T_i|Z_i)-n\e_n - \log \mu,
              \label{eq:bc-sc-10} 
\end{align}
where
(a) is from the constraints on the degrees of the graph,
(b) is obtained since $W_1$ and $W_2$ is a function of $S^n$ and
$T^n$,
(c) follows from Fano's inequality. 
Also, we can also bound the rate $R_1$ as
\begin{align}
  nR_1 &\geq H(W_1) \geq H(W_1|W_0)
              = I(S^n;W_1|W_0)+H(W_1|S^n,W_0) \\
              &\overset{(a)}{\geq} I(S^n;W_1|W_0)
              = H(S^n|W_0)-H(S^n|W_1,W_0)\\
              &\overset{(b)}{\geq} H(S^n|W_0)-n\e_n 
              = \sum_{i=1}^n
              H(S_i|S^{i-1},W_0)-n\e_n,\\
              &\overset{(c)}{\geq} \sum_{i=1}^n
              H(S_i|\underbrace{S^{i-1},T^{i-1},W_0}_{=Z_i})-n\e_n,\\
              &= \sum_{i=1}^n H(S_i|Z_i)-n\e_n, \label{eq:bc-sc-13}
\end{align}
where
(a) is obtained since $H(W_1|S^n,W_0) \geq 0$,
(b) follows from Fano's inequality, and 
(c) follows from adding conditioning.

Similarly, we also can obtain
\begin{align}
  nR_2 &\geq \sum_{i=1}^n H(T_i|Z_i) - n\e_n. \label{eq:bc-sc-14}
\end{align}

Therefore, using arguments similar to those of \cite{wyner75},
we can have the partial converse by dividing the
inequalities (\ref{eq:bc-sc-10}),
(\ref{eq:bc-sc-12}), (\ref{eq:bc-sc-13}), and (\ref{eq:bc-sc-14})
 by $n$, and taking the limit as $n\ra\infty$.
 \hfill $\blacksquare$

\end{document}